\documentclass[prb,aps,twocolumn,amsmath,amssymb,showpacs]{revtex4}
\usepackage{epsfig}
\psfigdriver{dvips}
\usepackage{graphicx}
\usepackage{bm,bbm}
\usepackage{dcolumn} 
\usepackage{times,color}

\usepackage{curves}
\usepackage{epic}
\usepackage{wasysym}
\usepackage{subfigure}

\def\beq{\begin{equation}}
\def\eeq{\end{equation}}
\def\bea{\begin{eqnarray}}
\def\eea{\end{eqnarray}}
\def\nn{\nonumber}

\def\nn{\nonumber}

\begin{document}

\title{       Valence bond crystal and possible
orbital pinball liquid in a $t_{2g}$ model
}

\author {     Fabien Trousselet }
\affiliation{ Max-Planck-Institut f\"ur Festk\"orperforschung,
              Heisenbergstrasse 1, D-70569 Stuttgart, Germany }

\author {     Arnaud Ralko }
\affiliation{ Institut N\'eel, UPR2940, CNRS et Universit\'e de Grenoble,
              Grenoble, F-38042 France}

\author {     Andrzej M. Ole\'s }
\affiliation{ Max-Planck-Institut f\"ur Festk\"orperforschung,
              Heisenbergstrasse 1, D-70569 Stuttgart, Germany }
\affiliation{ Marian Smoluchowski Institute of Physics, Jagellonian
              University, Reymonta 4, PL-30059 Krak\'ow, Poland }

\date{\today}

\begin{abstract}
We study a model for orbitally degenerate Mott insulators, where localized
electrons possess $t_{2g}$ degrees of freedom coupled by several, competing,
exchange mechanisms. We provide evidence for two distinct strongly fluctuating
regimes, depending on whether superexchange or direct exchange mechanism
predominates. In the superexchange-dominated regime, the ground state is
dimerized, with nearest neighbor orbital singlets covering the lattice.
By deriving an effective quantum dimer model and analyzing it numerically,
we characterize this dimerized phase as a valence bond crystal stabilized
by singlet resonances within a large unit cell.
In the opposite regime, with predominant direct exchange, the combined
analysis of the original model and another effective model
adapted to the local constraints shows that subleading perturbations
select a highly resonating ground state, with coexisting
diagonal and off-diagonal long-range orbital orders.

\textit{Published in: Phys. Rev. B} \textbf{86}, \textit{014432 (2012).}
\end{abstract}


\maketitle

\section{Introduction}
\label{intro}

One of the most intriguing concepts in contemporary condensed matter
physics is the possibility to stabilize phases without any broken
symmetry at low temperature. In strongly correlated materials, e.g. in
transition metal oxides, charge degrees of freedom are localized due to
large Coulomb interactions and both spin and orbital degrees of freedom
interact with each other and may remain disordered. While numerous
realizations of quantum spin liquids, such as resonating valence bond
(RVB) phases, have been reported both experimentally and theoretically,
\cite{Bal} there are only few examples of orbital liquids in Mott
insulators up to now.\cite{vdB04,G5,Zho10} Usually orbitals are accompanied
by spins and frustrated interactions in both subsystems could imply a
spin-orbital liquid phase. If spin and orbital variables are coupled to
each other by relativistic effects, such a phase may emerge
from effective pseudospin interactions on a honeycomb lattice.\cite{Cha10}
For the superexchange interactions in transition metal oxides,\cite{KK82}
a spin-orbital liquid could be stabilized in presence of geometrical
frustration. \cite{FiTs05} For instance, it was suggested as the ground state
on a triangular lattice for LiNiO$_2$ with active $e_g$ orbitals.\cite{VR6}
It would explain the absence of symmetry breaking with spin and orbital
order, but the realistic situation in this compound is more
subtle. \cite{Rei05} However, it was shown recently that the paradigm of a
spin-orbital liquid can be realized on the triangular lattice in a system with
active $t_{2g}$ orbitals,\cite{NO8,Nor11,CO11} given by the
$(111)$-plane of Ti$^{3+}$ ions in NaTiO$_2$.

The spin-orbital systems are rather complex and difficult to analyze as
spin and orbital operators are frequently
highly entangled \cite{Ole06} and it is not trivial to recognize which
part of the interactions is most frustrated. Therefore, considering
simpler and easier tractable models with only orbital variables (and
frozen spins) became recently fashionable as, in this way, the
intrinsically frustrated orbital interactions could be directly
investigated. These interactions have only the lattice symmetries,
\cite{vdB04} in contrast to high SU(2) symmetry of spin interactions.
For a two-dimensional (2D) $e_g$ orbital model one finds competing yet
robust orbital ordered phases.\cite{vdB99} In fact, in this case
frustration is not yet maximal and increases further when the
interactions are modified and the 2D compass model limit is approached.
\cite{Cin10} In this limit the system is still in the 2D Ising
universality class \cite{Nus08} and orbital order persists in a range
of finite temperature.\cite{Wen08}

Three-dimensional (3D) orbital models are characterized by an even
stronger frustration but an order by disorder mechanism stabilizes an
ordered phase in the $e_g$ orbital model,\cite{Nus04} as shown recently
by Monte Carlo simulations.\cite{Ryn10} The situation in the 3D
compass limit, also referred to as the classical $t_{2g}$ model,
is still controversial: while the absence of a phase transition at
finite temperature was conjectured by a high-temperature series
expansion,\cite{Oit11} a first order transition into a low-temperature
lattice-nematic phase without any orbital order was found by Monte
Carlo simulations.\cite{Wen11}

Geometrically frustrated $t_{2g}$ orbital systems on the triangular
lattice have been studied recently by effective models for large Coulomb
interactions.\cite{NO8,CO11,Nor11} These studies
led to identification of a variety of exotic ground states, depending
on the parameters governing electron hoppings and interactions.
Several of them were characterized
by dimerization into either spin- or orbital-singlet phases --- this
enables a description by an effective quantum dimer model (QDM),
\cite{RK88,Nor11,VR6} and opens a route towards the possibility of a
spin-orbital liquid\cite{CO11} in a regime with small Hund's exchange
coupling where both spin- and orbital degrees of freedom can fluctuate.
In a different situation where Hund's exchange coupling is large,
a ferromagnetic arrangement of spins is favored
--- here depending on which
exchange mechanism is dominant, either an orbital singlet phase, or a
non-dimerized phase characterized by local \textit{avoided-blocking}
constraints, were identified,\cite{NO8,CO11} but several questions
were left open about the precise nature of each phase.

The aim of this article is to understand the effects of the competition
between different exchange mechanisms naturally coexisting in such
systems, by studying a microscopic orbital model and deriving effective
Hamiltonians in extreme regimes.
In Sec.~\ref{ODM} we define the orbital model and analyze its low-energy
spectrum and some observables to identify three regimes with distinct
ground state properties. Then we analyze the regime where superexchange
dominates in Sec. \ref{supex}. This leads us to derive and analyze, in
Sec.~\ref{sec:qdm}, an effective QDM which allows us to characterize the
corresponding phase as an orbital valence bond crystal (VBC). Later on,
in Sec.~\ref{direx}, we focus on the opposite regime, where direct
exchange dominates, and identify by several approaches an exotic phase,
fluctuating and with long-range orbital order. Eventually we summarize
these findings in Sec.~\ref{ccl}.

\section{\label{ODM} Model and general approach}

\subsection{Model definition}

We investigate an orbital model on a triangular lattice, with exactly
one electron per site and $t_{2g}$ orbital degrees of freedom.
This model is a particular case of the one introduced in Ref.
\onlinecite{NO8} to describe the limit of strong Coulomb interactions
in layered, orbitally degenerate transition metal oxides, for instance
the Mott insulating titanate NaTiO$_2$.
In this compound, each titanium ion has a nominal valence $3+$,
corresponding to a single electron in the $3d$ shell; due to the
crystal field resulting from the octahedron formed by nearest neighbor
oxygens, this electron is confined to the $t_{2g}$ subspace
of this shell, characterized by the canonical basis:
$\{|yz\rangle, |xz\rangle, |xy\rangle \}$.
In the present layered structure we can consider a single $[111]$ plane of
the NaTiO$_2$ crystal, with $t_{2g}$ electrons described by a model
treating the combined effects of hoppings and Coulomb interactions on
a triangular lattice.

\begin{figure}[t!]
\begin{center}
\includegraphics[width=4.8cm]{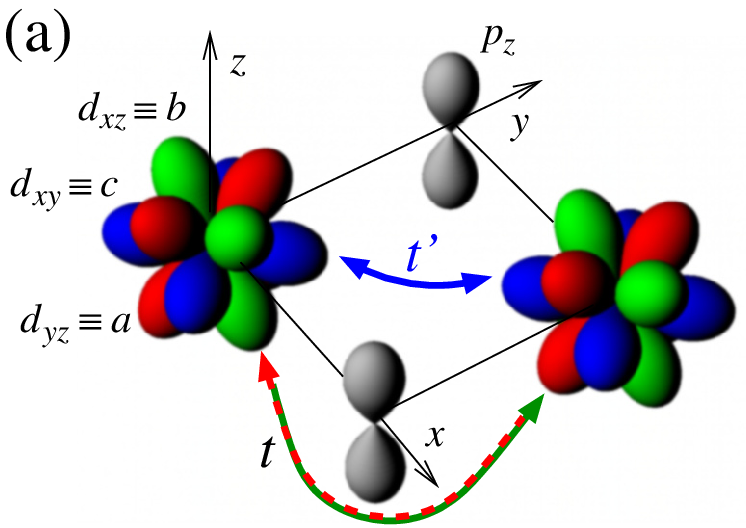}
\includegraphics[width=3.5cm]{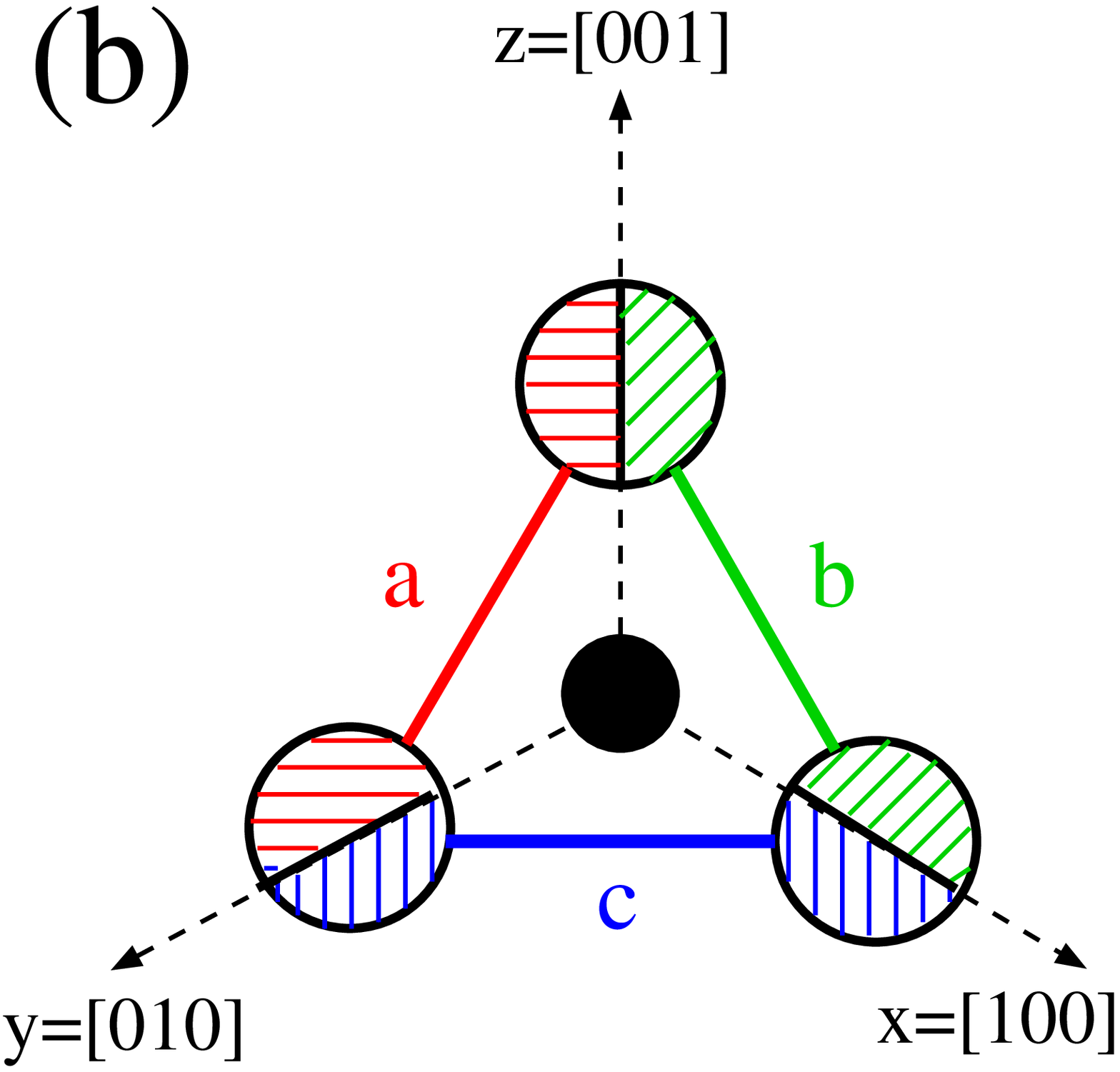}\\
\end{center}
\caption{\label{pzb} (Color online) (a) Representation of orbitals on
  two Ti ions, neighboring on a $c$-bond, and of the two distinct hopping
  processes between them: direct hopping of amplitude $t'$, and indirect
  hopping - via $p_z$ oxygen orbitals, with amplitude $t$. (b)
  An elementary triangle embedded in a $[111]$ plane; each bond of this
  triangle has a label  $\gamma \in \{a,b,c\}$
which labels both the bond direction and [as indicated in Fig.~\ref{pzb}(a)]
the orbital flavor active in direct exchange on this bond.}
\end{figure}

A general formulation of the electron dynamics and correlations requires
a multiorbital Hubbard model,\cite{KK82,NO8} taking into account all $t_{2g}$
orbitals and spin degrees of freedom.
For convenience, in the following we label each bond direction with an index
$\gamma \in\{a,b,c\}$ corresponding to the plane (respectively $yz$, $xz$ and
$xy$) in which the bond is embedded. Kinetic terms are of two types:
(i) direct hopping between ions neighboring on the triangular lattice, with
amplitude $t'$ --- there, the electron
can hop only between orbitals of a single flavor, depending on the
direction of the bond [this flavor is e.g. $xy \equiv c$ for a bond in the
$xy$ plane of the cubic lattice, see Fig.~\ref{pzb}(a,b)] ---
in consequence we rename the orbital flavor allowing for direct hopping
in this direction with the same index,\cite{G5} that is
\begin{equation}
|a\rangle \equiv |yz\rangle, \hskip .7cm
|b\rangle \equiv |xz\rangle, \hskip .7cm
|c\rangle \equiv |xy\rangle;
\end{equation}
(ii) indirect hopping with amplitude $t$, where
an electron hops, from a Ti ion, first to a neighboring oxygen and then
to another Ti ion, neighbor of the former on the triangular lattice.
This latter hopping process involves exclusively the two orbital flavors
not involved in direct hopping on this bond; more precisely, considering
two ions neighboring on a $c$ bond, an electron initially in the $a$
orbital of one ion can hop with indirect hopping only to the $b$ orbital
of the other ion, and vice-versa [see Fig.~\ref{pzb}(a)]. Independently,
the on-site Coulomb interactions are governed by Hubbard and Hund's
exchange parameters $U$ and $J_H$, respectively.

In the limit where these interaction amplitudes are larger than hopping
parameters, the $3d$ electrons localize and the system is Mott insulating.
In this regime of parameters one can use an effective
model acting within the subspace with one electron per site
(in the present case of $d^1$ configuration of Ti$^{3+}$ ions),
which describes interactions between their
spin and orbital degrees of freedom. Such
a model was derived for layered titanates in a previous work, see
Ref.~\onlinecite{NO8}, and contains exchange couplings. These follow
from second-order processes where one electron hops from a site (Ti ion)
to a neighboring site which becomes doubly occupied at an energy cost
given by a linear combination of $U$ and $J_H$, before
one of the two electrons hops to the site left empty. The coexistence of
direct- and indirect- hoppings implies that three exchange mechanisms
are allowed. We list them below, indicating how their amplitudes depend
on the hopping parameters $t$, $t'$:
\newline\noindent
(i) direct-exchange terms which result from processes where both hopping
processes involved are direct hoppings, and are thus of amplitude
$\propto t'^2$;
\newline\noindent
(ii) superexchange terms resulting from processes where both hoppings
involved are indirect hoppings, and have an amplitude $\propto t^2$; and
finally
\newline\noindent
(iii) mixed-exchange terms resulting from processes that combine one
direct hopping and one indirect hopping, and their amplitude is thus
$\propto t t'$.
\newline\noindent
These exchange terms couple both spin and orbital degrees of freedom
of nearest neighbor (n.n.) ions on a bond $\langle ij\rangle$.
As a consequence of the $SU(2)$-symmetry in spin space of the original
Hubbard model, each of them can be written as the sum of a term
with a projection operator $(\vec{S}_i\cdot \vec{S}_j-\frac14)$ on a spin
singlet state on the bond $\langle ij\rangle$, and a term with a
projection operator $(\vec{S}_i\cdot \vec{S}_j+\frac{3}{4})$, selecting
a triplet state for this bond.

We now specify further the context to a situation with large Hund's
exchange coupling $J_H$ (i.e., close to its maximal allowed value $U/3$),
when only spin triplet states contribute to exchange processes.
This favors the alignment of spins into a ferromagnetic phase, where
exchange terms act only on orbital degrees of freedom. This situation
can actually also be realized, even if $J_H$ alone is not large enough
to polarize spins, in the presence of an external magnetic field, the
effect of which adds to that of Hund's exchange --- for NaTiO$_2$ the
estimated value of $\eta=J_H/U$ is $\eta\simeq 0.14$, i.e., not far
from the value of the estimated transition to the spin-polarized phase
$\eta_c \simeq 0.16$.\cite{CO11} In such
a spin-polarized phase, setting $J=(t^2+t'^2)/(U-3J_H)$ as the unit of
energy ($J=1$), the effective exchange Hamiltonian for orbital degrees
of freedom is:
\begin{eqnarray}
\label{Hfm}
{\cal H}  = (1-\alpha) H_{\rm s} + \sqrt{\alpha(1-\alpha)}
H_{\rm m} + \alpha H_{\rm d}\,,
\end{eqnarray}
where $H_{\rm s}$, $H_{\rm d}$ and $H_{\rm m}$ are, respectively, the
superexchange, the direct exchange and the mixed exchange Hamiltonians
defined below. The parameter
\begin{equation}
\alpha=\frac{t'^2}{t^2+t'^2}
\end{equation}
interpolates continuously between the superexchange
($H_{\rm s}$, for $\alpha=0$) and  direct exchange ($H_{\rm d}$, for
$\alpha=1$) Hamiltonians --- while the additional mixed-exchange term
$H_{\rm m}$ is present only for $0<\alpha<1$, in combination with the
others two. The three terms of Eq.~(\ref{Hfm}) are given by:
\begin{eqnarray}
\label{Hs}
H_{\rm s}&=& 2\sum_{\langle ij \rangle \parallel \gamma}\!\Big\{
\left(T_{i \gamma}^{+} T_{j \gamma}^{+} + {\rm c.c.}\right)
+ \sum_{\mu \ne \gamma} n_{i \mu} n_{j \bar{\mu \gamma}} \Big\},
\\
\label{Hm}
H_{\rm m}&=& -\sum_{\langle ij \rangle \parallel \gamma}
\sum_{\mu \ne \gamma} \left(T_{i\mu}^+ T_{j\bar{\mu \gamma}}^+ +
           {\rm H.c.} \right)\, ,\\
\label{Hd}
H_{\rm d}&=& 2 \sum_{\langle ij \rangle \parallel \gamma}
n_{i\gamma} n_{j\gamma} .
\end{eqnarray}
The electron creation operator in orbital $\gamma=a,b,c$ at site $i$
is $\gamma_i^{\dagger}$, $n_{i\gamma}\equiv\gamma_i^\dagger\gamma_i^{}$,
and the single-occupancy local constraint reads
$\sum_\gamma n_{i\gamma}=1$. The notation
$\langle ij\rangle\parallel\gamma$ indicates that a bond between sites
$i$ and $j$ is oriented along the lattice direction $\gamma$,
see Fig.~\ref{pzb}(b); in other words, this bond is parallel to the
vector $\vec{e}_\gamma$ defined by $\vec{e}_c=(1,0)$,
$\vec{e}_{a/b}=(\pm 1/2, \sqrt{3}/2)$
[see Figs. \ref{clust}(a) and \ref{clust}(b)]. In the expressions of
super- and mixed exchange terms we use pseudospin operators
$\vec{T}_{i \gamma}$ defined for each site \textit{and}
flavor index; for instance for $\gamma=c$:
\begin{eqnarray}
\label{Tz}
T^z_{ic}&\equiv& \frac12\left(n_{ia}-n_{ib}\right),\\
T^{+}_{ic}&\equiv& a^{+}_{i}b_i^{} = (T^-_{i c})^\dagger.
\end{eqnarray}
The operators $\vec{T}_{ia}$
and $\vec{T}_{ib}$ are obtained by cyclic permutation of flavor indices.
The orbital flavor active in direct exchange for each bond direction is
depicted in Fig.~\ref{pzb}(b); in superexchange on the same bond, both
other flavors (and only those) are active. In Eqs.~(\ref{Hs}) and (\ref{Hm})
$\bar{\mu\gamma}$ is the flavor index distinct from both $\mu$ and $\gamma$.

\subsection{\label{numal} Numerical analysis of the low-energy spectrum}

\begin{figure}[t!]
\begin{center}
\includegraphics[width=4.4cm]{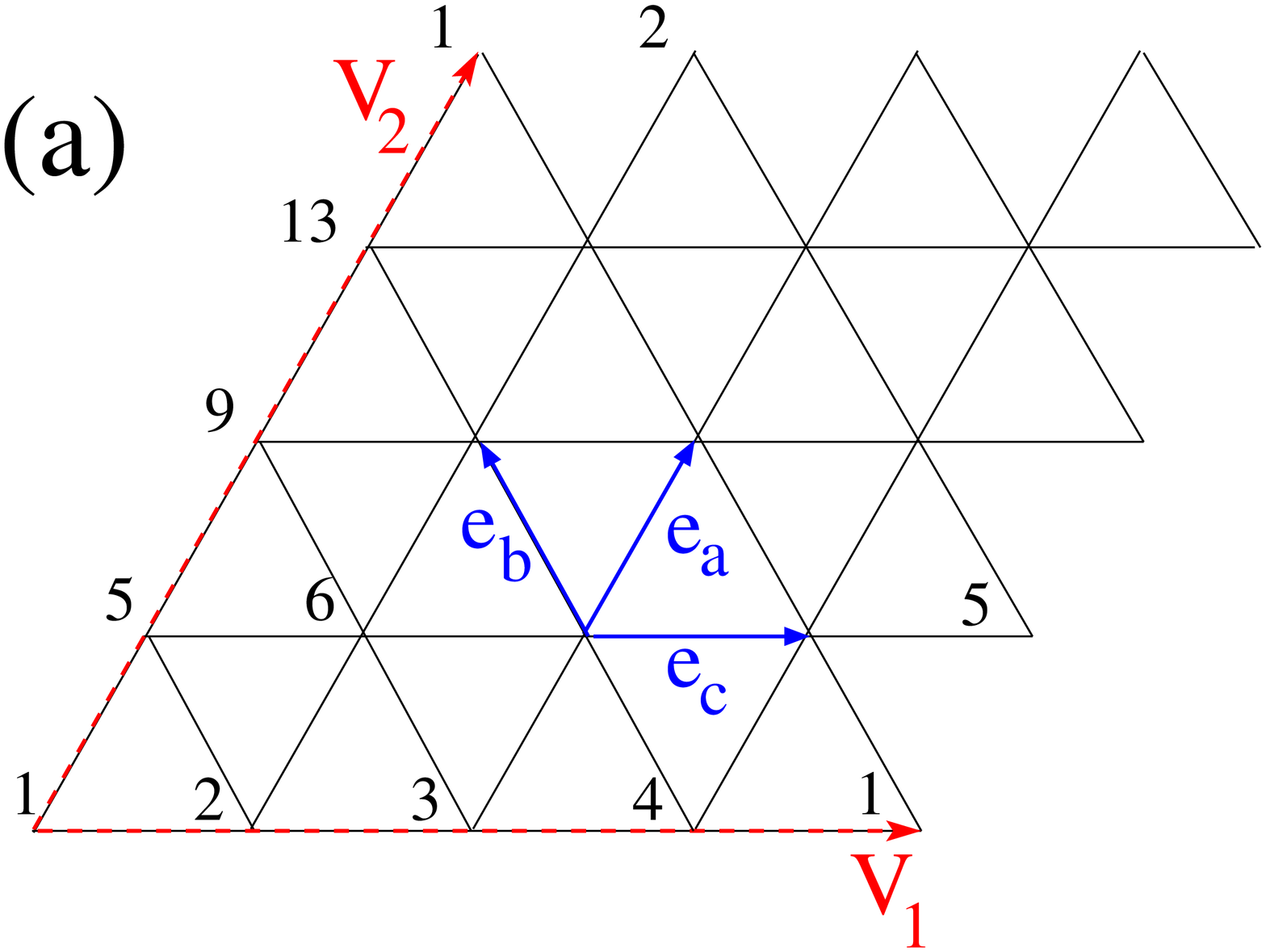}
\includegraphics[width=3.6cm]{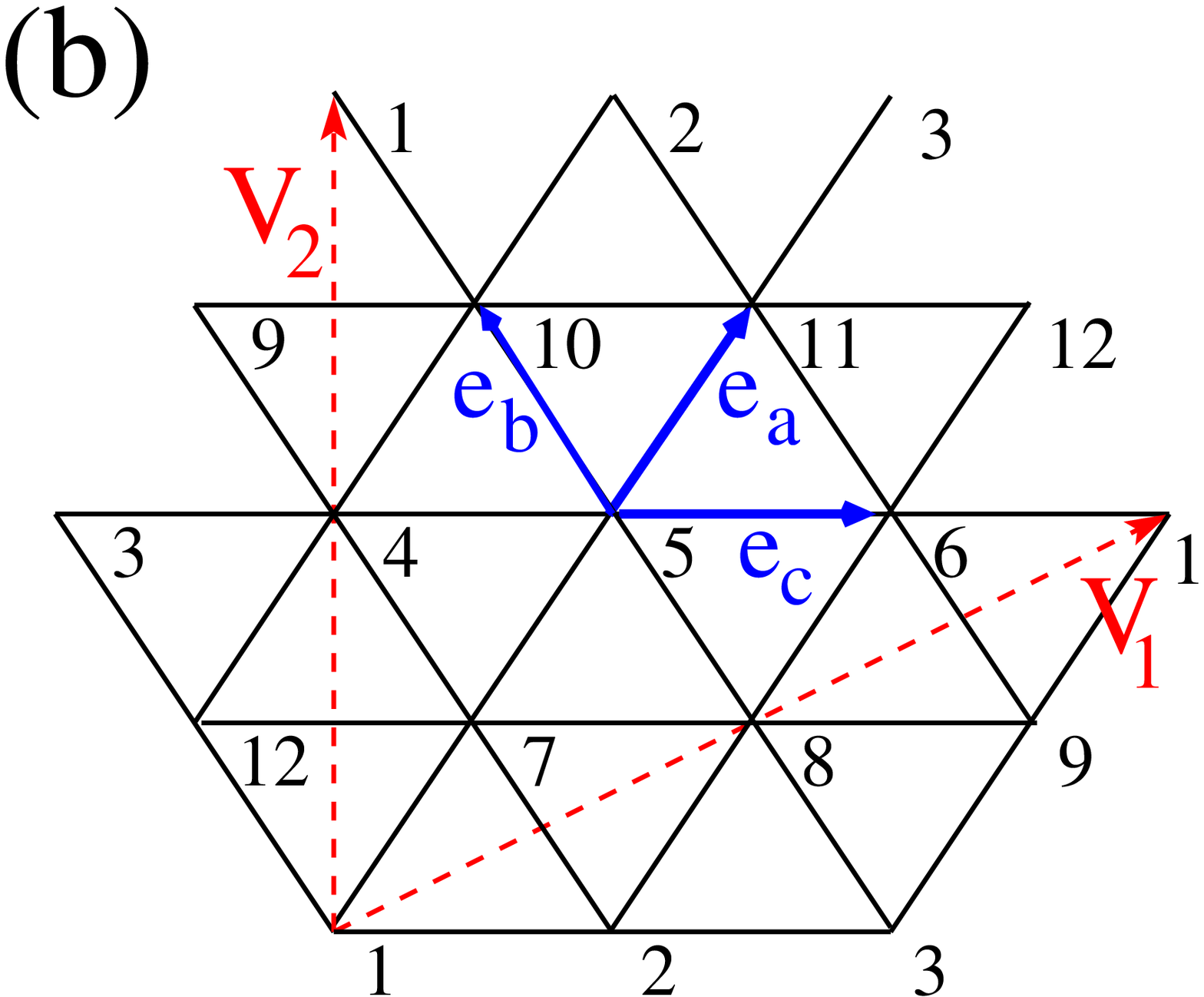}\\
\includegraphics[width=3.8cm]{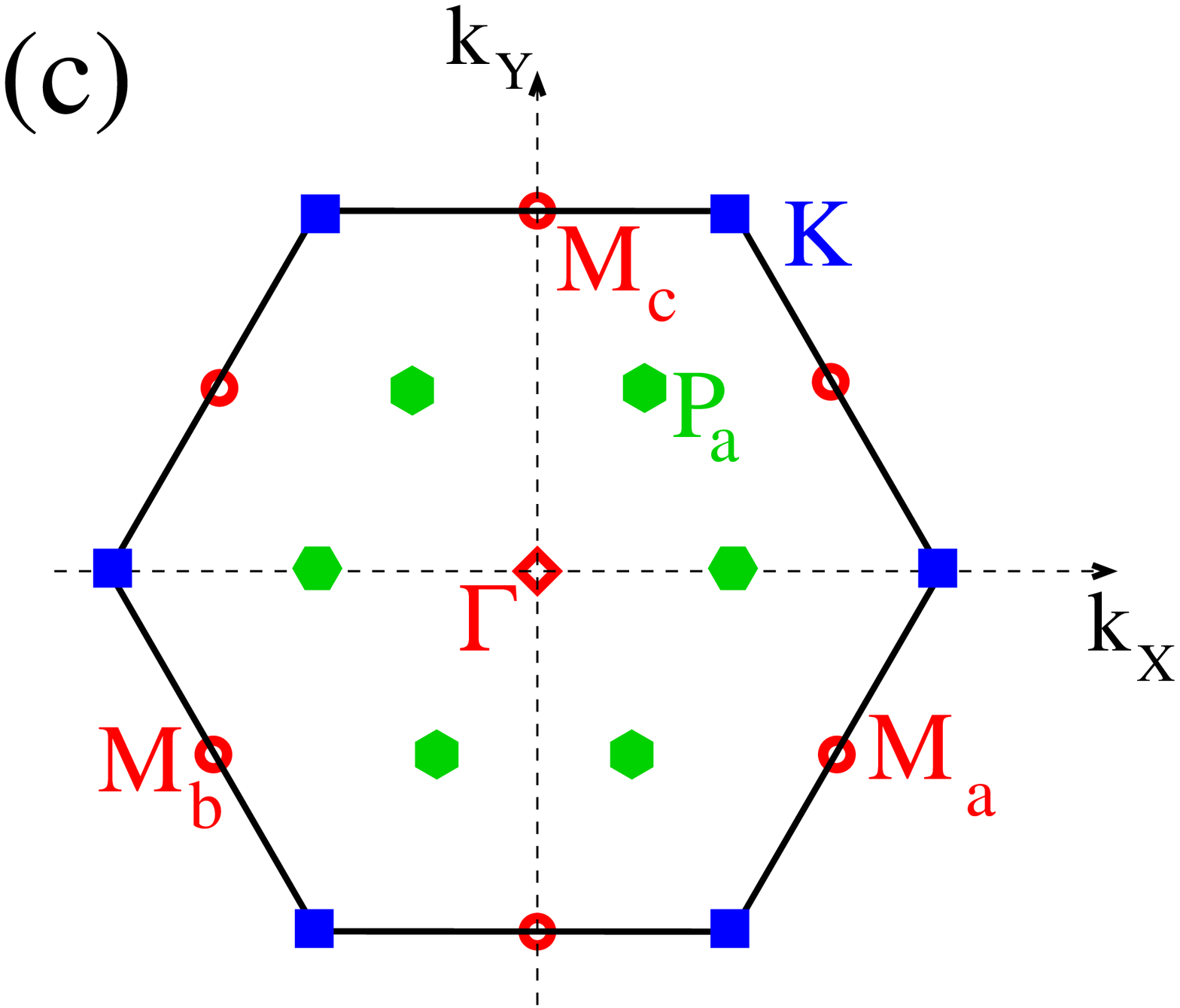}\\
\end{center}
\caption{\label{clust} (Color online)
Clusters used for the numerical study of the orbital Hamiltonian
Eq.~(\ref{Hfm}), with: (a) $N=16$ sites and (b) $N=12$ sites.
The multiple occurrence of site indices indicates the periodic boundary
conditions, i.e., invariance under translations by vectors
$\vec{V}_{1/2}$. (c) First Brillouin zone of the triangular lattice,
with $\Gamma$, ${K}=2{P}_a$ and ${M}_{a/b/c}$ points shown ---
$M_c=(0,2\pi/\sqrt{3})$ in terms of $(k_x,k_y)$ coordinates.}
\end{figure}

The Hamiltonian $\cal{H}$ [Eq.~(\ref{Hfm})] is studied by L\'{a}nczos exact
diagonalization (ED) on periodic
clusters. Most generally, these are characterized by two
(linearly independent) vectors $\vec{V}_1$ and $\vec{V}_2$: those with
the smallest possible value of $|\vec{V}_1 \times \vec{V}_2|$
such that the image of any site by either of these vectors is identified
to this site by periodicity. We choose to consider clusters invariant
under point group symmetries of the triangular lattice; for this, the
simplest choice of vectors $\vec{V}_i$ is e.g. $\vec{V}_1=L.\vec{e}_c$,
$\vec{V}_2=L.\vec{e}_a$, which corresponds to a $N=L^2$-site cluster as
in the case $N=16$, see Fig.~\ref{clust}(a). Another possible choice is
to take $\vec{V}_{1/2}=L.(\vec{e}_a+\vec{e}_{c/b})$; the corresponding
cluster has $N=3L^2$ sites; we will use
hereafter the $N=12$ cluster shown in Fig.~\ref{clust}(b)
(and similar but larger clusters in Sec.~\ref{sec:qdm}).

\begin{figure}
\begin{center}
\includegraphics[width=0.48\textwidth,clip]{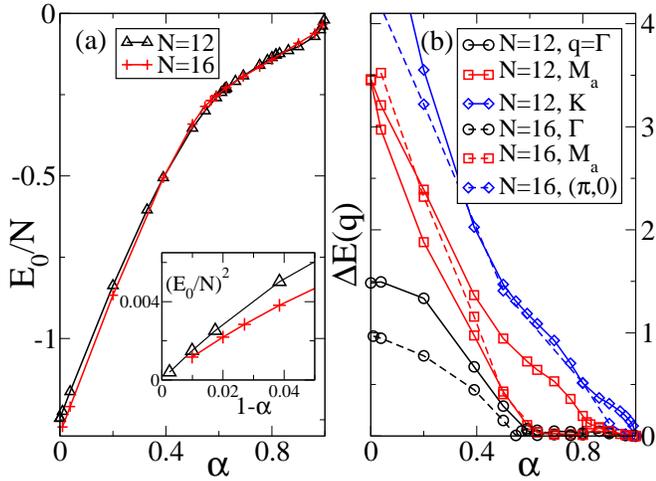}\\
\caption{\label{nrj}(Color online)
(a) Ground state energy per site $E_0/N$ as function of $\alpha$ for periodic
clusters of $N=12$ and $16$ sites. Inset: square of the previous
quantity as function of $(1-\alpha)$ in the vicinity of $\alpha=1$.
(b) Lowest excitation energies $\Delta E(\vec q)$ for momenta
$\vec q=\Gamma$, $M_a$, and (depending on cluster size) either
$K$ or $(\pi,0)$, see Fig. \ref{clust}(c).}
\end{center}
\end{figure}

The momentum-resolved low-energy spectrum gives the energy $E_0$ of the
ground state, found at the $\Gamma$ point. The $\alpha$ dependence of
$E_0$, see Fig.~\ref{nrj}(a), shows a striking contrast between the two
opposite regimes:
(i) $\alpha\le 0.6(1)$ where $E_0$ increases linearly with $\alpha$, and
(ii) $\alpha$ close to 1, where the $\alpha$-dependence of $E_0$ is
$\propto\sqrt{1-\alpha}$
[see inset in Fig.~\ref{nrj}(a)]. The latter scaling indicates that the
low-energy dynamics is here dominated by mixed-exchange terms.

Considering low-energy excitations shown in Fig.~\ref{nrj}(b), one notices an
intermediate regime (iii) for $0.6 \lesssim \alpha \lesssim 0.8$, where
several quasi-degenerate lowest states at $\Gamma$ and $M_\gamma$ points
are well separated from higher excitations. We checked that there are
exactly six such low-energy states. Three of them, at the $\Gamma$ point,
consist of a pair of exactly degenerate states, which are the exact ground
states for a finite range of $\alpha$ in this regime,\cite{3sym} and a
third low-energy state which is the lowest otherwise. The boundaries
$\alpha_{cr,\pm}$ of the zone with exact ground state degeneracy
depend slightly on cluster size, e.g. the upper boundary
is $\alpha_{cr,+}\simeq 0.81(1)$ and $0.72(1)$ for $N=12$ and $N=16$
respectively. The three remaining low-energy states are degenerate, one being
at each $M_\gamma$ point. These features indicate a gapped, ordered phase with
six-fold degeneracy of the ground state in the thermodynamic limit (TL).

In contrast, in regime (ii), although lowest states are also found at
 $\Gamma$ and $M_\gamma$ points indicating a symmetry-broken phase,
above them, excited states of energy $\sim c_{N,\vec{q}}\sqrt{1-\alpha}$
are compatible with modes becoming gapless in the TL --- for details see
discussion in Sec.~\ref{direx}. Eventually, for $0\le\alpha\lesssim 0.6(1)$,
the lowest excitations are at the $\Gamma$ point and have energies $O(1)$
decreasing with increasing $N$, compatible with an order breaking
non-translational symmetries. Note, however, that since both clusters
considered are of moderate size and with different shapes, we cannot
characterize this phase from their low-energy spectra alone.

\section{Dimerization in the superexchange regime}
\label{supex}

In this Section we focus on the regime where superexchange is the
dominant exchange mechanism
(i.e., we consider the regime of $\alpha \ll 1$). We will
provide, within the orbital model Eq.~(\ref{Hfm}), evidence for a
dimerized ground state, which we will study further in the next Section
using an effective model.

\begin{figure}[t!]
\begin{center}
\includegraphics[width=7.8cm]{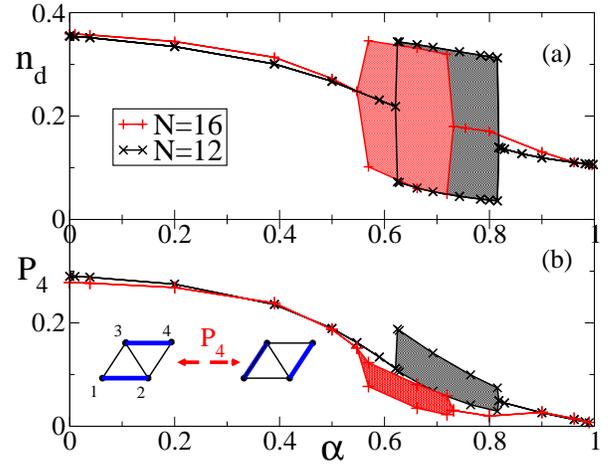}\\
\end{center}
\caption{\label{ndim}(Color online)
(a) Dimer occupation number $n_d(c)$, and (b) the resonance amplitude $P_4$
[Eq.~(\ref{P4}), with site indices as indicated on the left-hand-side lozenge;
each lozenge in the inset shows a singlet covering at play in the resonance]
for $N=12$ and $16$ as functions of $\alpha$. Shaded areas
indicate the range of possible values due to an exact twofold ground state
degeneracy occurring for both clusters over a finite range of $\alpha$ -
see text and Ref.~\onlinecite{CO11}.}
\end{figure}

The superexchange interactions alone, in the present context, are known
to favor dimerization into orbital singlets on nearest neighbor bonds.
\cite{NO8} This is clear when one considers an isolated bond $\langle
ij \rangle \parallel \gamma$, for which the ground state of $H_{\rm s}$
is a singlet wave function in terms of the pseudospin variables
$\vec{T}_{i,\gamma}$ and $\vec{T}_{j,\gamma}$. On larger systems, we will
evaluate the strength of dimerization by computing the dimer or singlet
expectation value, defined as:
\beq
n_d=\left\langle\Psi_0\left| d^\dagger_{ij}d_{ij}^{}\right|\Psi_0\right\rangle,
\label{eqnd}
\eeq
where $|\Psi_0\rangle$ is the ground state found in ED, and
\beq
d^\dagger_{ij}\equiv\frac{1}{\sqrt{2}}
\left(a^\dagger_ia^\dagger_j-b^\dagger_ib^\dagger_j\right)
\eeq
is the operator creating a singlet from the electron vacuum for a
$c$-bond $\langle ij\rangle\parallel c$. While for $\alpha\rightarrow 1$
one finds rather small values of $n_d \simeq 1/9$, see Fig.~\ref{ndim}(a),
a much larger value $\simeq 0.36(2)$ is found for $\alpha \ll 1$,
indicating a dimerized ground state.
Note that the value $n_d \simeq 1/9$ is close to the values expected for
uncorrelated orbitals and almost unaffected by the avoided-blocking
constraints. Note as well that the shaded areas in Fig.~\ref{ndim} for
$0.6 \lesssim \alpha \lesssim 0.8$ are \textit{not}
resulting from computational uncertainty, but from the exact ground state
degeneracy occurring in the corresponding (size-dependent) range
of parameters. Depending on the actual ground state $|\Psi_0\rangle$ selected
within the ground state manifold (here a 2D space),
for an observable $\cal{O}$ not commuting with $\cal{H}$,
$\langle \Psi_0 |\cal{O}$ $|\Psi_0 \rangle$ can take all possible values
in the range indicated by the shaded area (and delimited by the same
symbols as in the non-degenerate case).

Interestingly, the value of $n_d$ obtained for $\alpha \ll 1$ is close to
$7/24$ obtained for a variational state $|\Psi_{\rm var}\rangle$, built as an
equal-amplitude superposition of the twelve columnar singlet coverings which
minimize the energy $\langle \Psi_{\rm var}| H_{\rm s}|\Psi_{\rm var} \rangle$
among static singlet coverings. However, the true ground state has a much lower
energy and singlet correlations significantly less modulated than those
estimated with $|\Psi_{\rm var}\rangle$, see Appendix A. Hence
we compute the quantity:
\beq
P_{4}=\left\langle \Psi_0\left|(d^\dagger_{13} d^\dagger_{24}d_{12}d_{34} +
{\rm H.c.})\right|\Psi_0\right\rangle,
\label{P4}
\eeq
defined on a lozenge and shown as function of $\alpha$, see Fig.~\ref{ndim}(b).
A discussion of the value of $P_4$ on an isolated lozenge and its
interpretation in terms of singlet resonance are given in Appendix A.
The large values of $P_4$ shown in Fig.~\ref{ndim}(b) for the dimerized
ground state of the superexchange regime suggest that it is
stabilized by such resonances between nearest neighbor singlets. These
resonances can favor either a valence bond crystal (VBC) or a dimer
liquid; in the first case large quantum fluctuations may
reduce strongly the amplitude of any order parameter related to this order, 
in comparison with the amplitude expected for a model VBC wave function
(such as $|\Psi_{\rm var}\rangle$ for a columnar phase). Thus, the
identification of the phase cannot be
addressed directly in ED within the orbital model Eq.~(\ref{Hfm}), but
requires to use an adapted effective model, which we describe hereafter.

\section{Quantum Dimer Model}
\label{sec:qdm}

\subsection{\label{derqdm} Derivation of a Quantum Dimer Model}

Motivated by the evidence discussed above for a dimerized phase in the
superexchange regime, we derive a quantum dimer model (QDM) to provide
a better understanding of this phase. Such a model, thanks to the
reduced number of degrees of freedom at equal size (here the degrees of
freedom are not anymore orbital configurations at
each site, but the positions of dimers in close-packed dimer coverings
fulfilling a hard-core constraint), can be accessed numerically for
much larger system sizes than the original orbital model of
Eq.~(\ref{Hfm}). It is then possible to address the issue of the
behavior in the TL. For this purpose, we follow the Rokhsar-Kivelson (RK)
scheme;\cite{RK88,Nor11} in the present context this means that we project
the orbital Hamiltonian in the subspace spanned by nearest neighbor orbital
singlet coverings
$|\cal{C}\rangle$. An important characteristic of these
coverings is that they form an overcomplete, thus non-orthogonal basis. Thus,
instead of diagonalizing the Hamiltonian projected onto the singlet coverings'
subspace (with matrix elements $\langle \cal{C}'|\cal{H}|\cal{C}\rangle$)
one has to solve a generalized eigenvalue problem involving both this
matrix and an
overlap matrix with elements $\langle \cal{C}'|\cal{C}\rangle$. Due to the
high connectivity of these matrices the numerical solution of this problem is,
a priori, much more time consuming than the diagonalization of a sparse matrix
of the same size.

An approximation done in similar cases \cite{RK88}
consists of expanding the matrix elements mentioned above in powers of a
parameter $x$ - here, this parameter is defined formally such that the
singlet wave function on a bond $\langle ij\rangle\parallel c$ is
$x(|a_i a_j\rangle - |b_i b_j\rangle)$ (similar expressions hold for singlets
on bonds oriented along either $\vec{e}_a$ or $\vec{e}_b$). Note that,
although $x$ takes the value $1/\sqrt{2}$ in the case of interest, here it is
introduced as a control parameter for the perturbative expansion. An effective
Hamiltonian matrix, in an orthogonal basis of dimer coverings $|c\rangle$
which are in one-to-one correspondence with singlet coverings
$|\cal{C}\rangle$, is then obtained at given order $p$ of this expansion.

\begin{figure}[t!]
\begin{center}
\includegraphics[width=8.0cm]{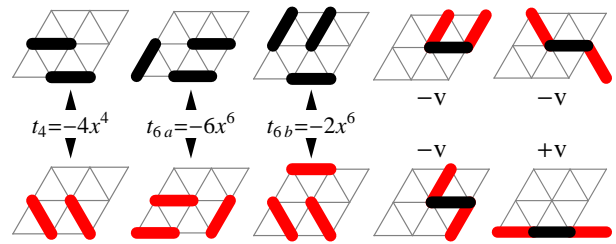}\\
\caption{\label{loopot} (Color online)
  Kinetic processes (three left columns) and configurations
  with finite potential energies
  $\lambda_c=\pm v$ (two right columns) in the effective QDM
  Hamiltonian $\cal{H}_{\rm QDM}$ of Eq.~(\ref{qdm})
derived in the superexchange regime.
The coupling between configurations in the leftmost column, with amplitude
$t_4$, is found at 4$^{th}$ order in overlap expansion, with $x=1/\sqrt{2}$;
subsequent columns show couplings on two inequivalent six-bond loops,
with amplitudes $t_{6a}$ and $t_{6b}$ found at sixth order in this expansion.
  A potential term with amplitudes $\pm v$ applies to every
  interdimer bond when dimers touching this bond are parallel to each
  other (the sign depends on whether they are parallel or not to the
  interdimer bond).}
\end{center}
\end{figure}

We carried out this derivation up to order $p=6$, and obtained the following
effective Hamiltonian:
\begin{equation}
\label{qdm}
{\cal H}_{\textrm{QDM}} = \sum_c \lambda_c | c \rangle \langle c |
- \sum_{c,c'} t_{c,c'} |c \rangle \langle c'|\,,
\end{equation}
where dimer coverings $|c \rangle$ and $| c'\rangle$ differ by a shifted
closed loop of length $l=4$ or $6$. The off-diagonal terms of amplitudes
$t_{c,c'}$ and diagonal terms of amplitudes $\lambda_c$ are explicited below
and schematized in Fig. \ref{loopot}; these amplitudes are expressed in units
of the superexchange amplitude $J(1-\alpha)$.

\begin{itemize}
\item[(a)]
Off-diagonal (kinetic) terms shown in three first columns of
Fig.~\ref{loopot} --- each term
$\propto |c\rangle\langle c'|$ couples all pairs of dimer configurations
which differ from each other \textit{only} by the position of
dimers on a short loop of length $l \le 6$. These terms originate from
the off-diagonal part of superexchange interactions on interdimer bonds.
The term with amplitude $t_4=-4x^4$, which flips two dimers on a $l=4$-long
loop, is found at fourth order in the derivation; at sixth order, two terms
flipping three dimers on loops of length $l=6$ have amplitudes
$t_{6a}=3t_4/4$ and $t_{6b}=t_4/4$, depending on the shape of the loop.
\item[(b)]
Diagonal terms $\propto |c\rangle\langle c|$, which play a role of
effective potential terms (found at order 4), act on interdimer bonds
and favor certain energetically optimal states. These terms
originate from the diagonal part of superexchange terms on these bonds.
The resulting potential energy $\lambda_c$ can be written (after a
global energy shift) as a sum over interdimer bonds of terms
taking values $\pm v$ when both dimers touching a bond are parallel
to each other (see Fig.~\ref{loopot}), and $0$ otherwise. In the present
derivation we have found that $v=t_4/2$.
\end{itemize}

This effective QDM mimics the accurate properties of the original model
in this regime, similarly as in Refs.~\onlinecite{VR6} and
\onlinecite{RMP} for spin-singlet phases. Note that the potential energy
is different from the one of the RK QDM on the triangular lattice
\cite{3MS} (where it is proportional to the
number of lozenges with dimers occupying two parallel edges);
here the potential acts on each interdimer bond instead of lozenges.

Although we derived ${\cal H}_{\textrm{QDM}}$ for the purpose of studying the
superexchange limit $\alpha=0$, it is also relevant in the regime of
$0<\alpha \ll 1$
where dominant superexchange couplings are perturbed by small mixed-exchange
couplings (with amplitude $\simeq \sqrt{\alpha}$). Indeed, matrix elements of
the latter do not connect different singlet coverings, thus they do not modify
the effective QDM (the direct-exchange perturbation bring some potential terms
which, if taken into account, would modify $\cal{H}_{\textrm{QDM}}$, but
should be of amplitude $\propto \alpha$ and thus negligible in the limit
considered). The main
effects of mixed- and direct-exchange terms are to contribute to
quantum fluctuations out of the n.n. singlet covering subspace, and to
destabilize the dimerized phase for increasing $\alpha$.

\begin{figure}[t!]
\begin{center}
\includegraphics[width=3.9cm]{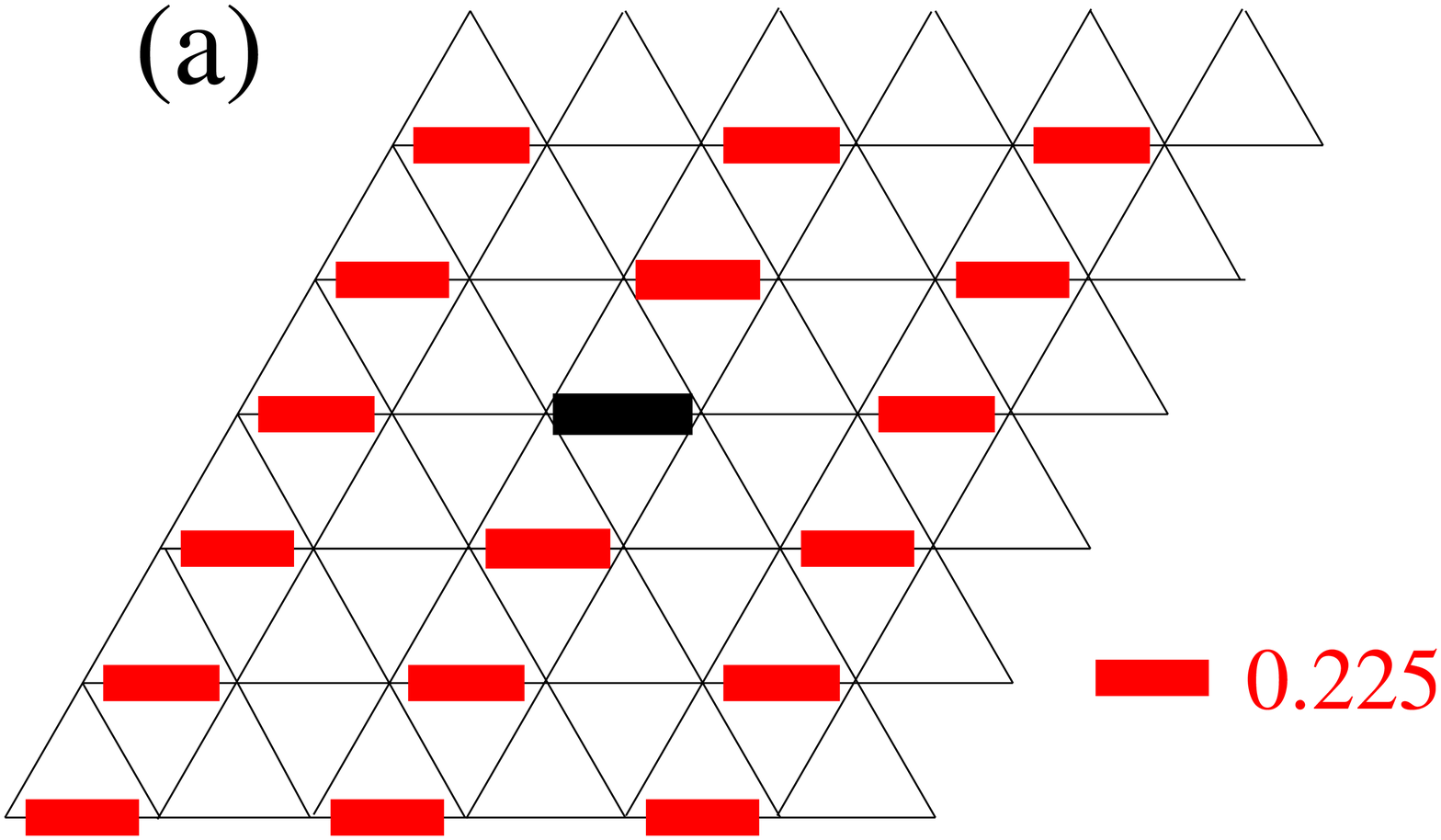}
\hspace{0.2cm}
\includegraphics[width=4.0cm]{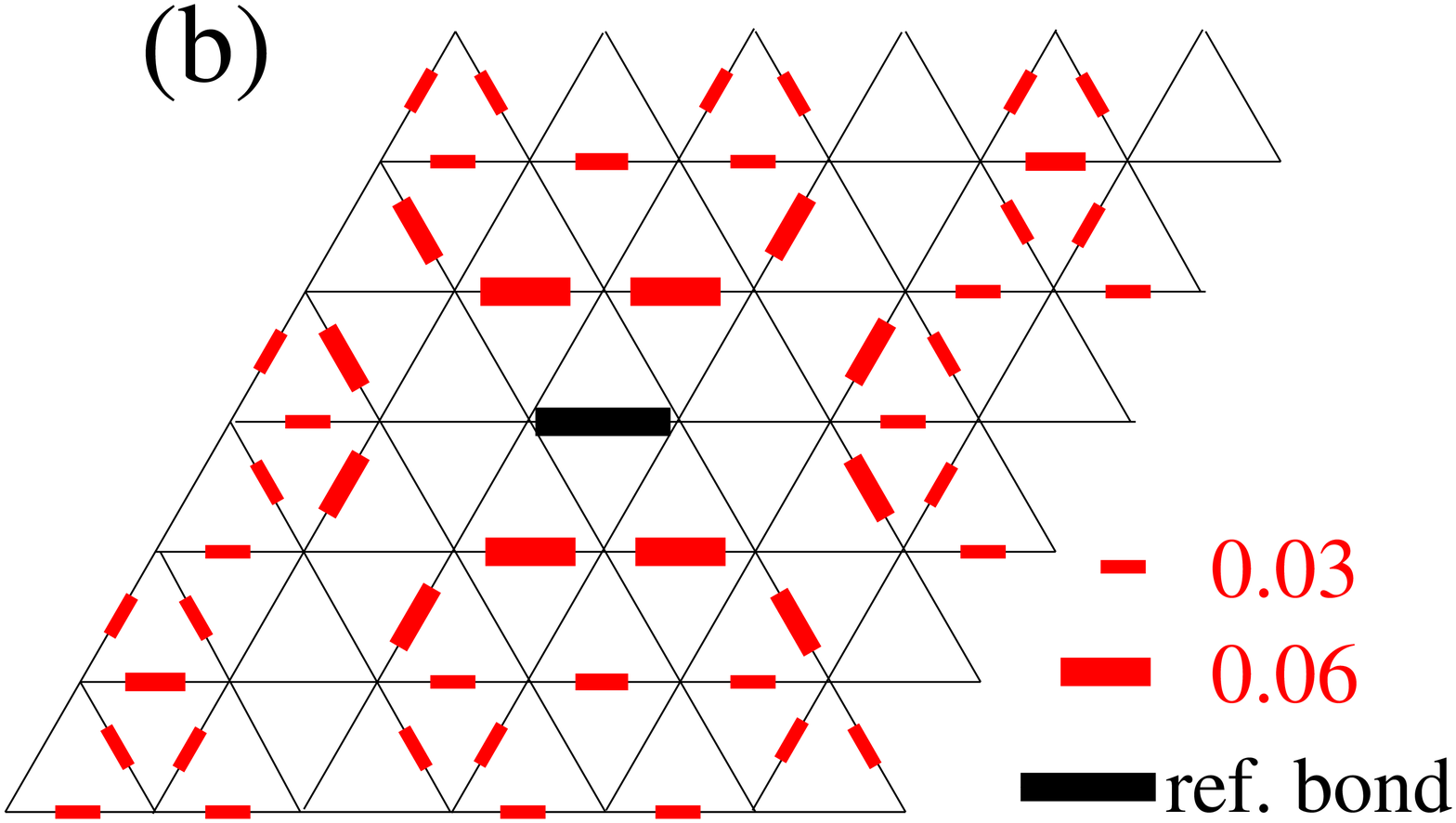}\\
\vspace{0.4cm}
\includegraphics[width=7.7cm]{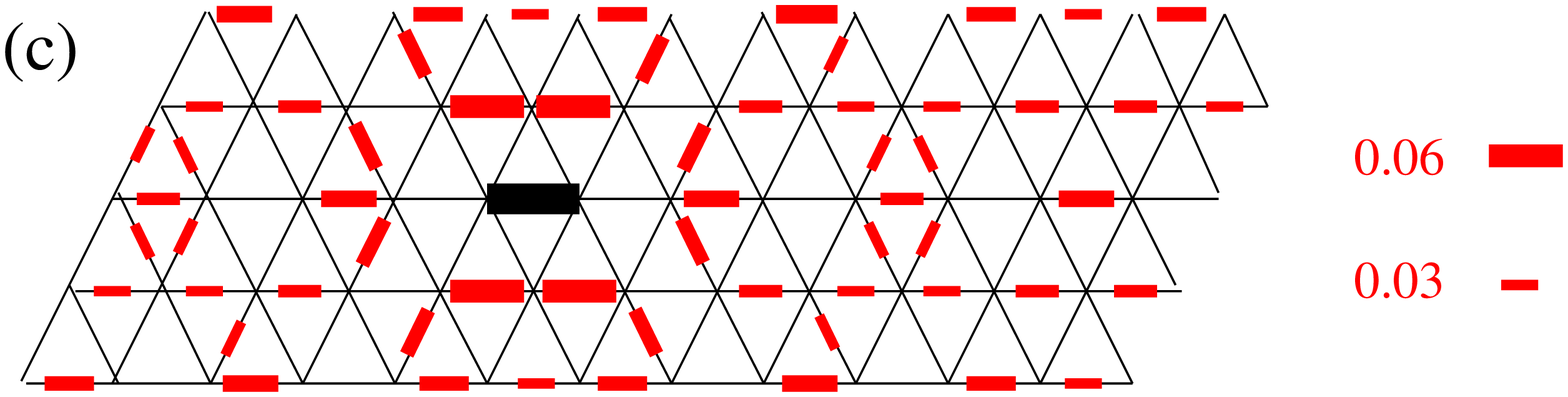}\\
\caption{\label{corrd} (Color online) Real-space dimer correlations obtained
  for the QDM Eq. (\ref{qdm}) on periodic clusters: (a,b) on the $N=36$
  cluster, for (a) $v=1.5$ and (b) $v=0.5$;
and (c) on the $N=48$ cluster
 for $v=0.5$. In (a) correlations correspond to the ground state of the
$(1,1)$ topological sector only, see Sec.~\ref{tgap}. }
\end{center}
\end{figure}

\subsection{\label{studm} Analysis of the quantum dimer model}

We have studied ${\cal H}_{\textrm{QDM}}$ using two numerical methods:
(i) ED on periodic clusters with $N=12$, $36$ and $48$ sites --- for the
largest one the use of translational symmetries and topological invariants
(see Sec.~\ref{tgap}) allows us to reduce the Hilbert space size to
$N_H\simeq 1.0\times 10^7$ in a representation of momentum $\Gamma$; and
(ii) a zero temperature Green's function quantum Monte Carlo (GFMC).\cite{GFMC}
This method can be performed here since all the off-diagonal terms
are negative. It allows us to obtain the ground states of significantly larger
clusters than within ED; here we focus on periodic clusters of $N=3L^2$ sites up
to $L=10$. We choose to set, unless they are explicitly specified, the amplitudes
of off-diagonal terms to their values obtained in the derivation, and adopt
$t_4$ as unit of energy in this section. But we keep $v$ as a free
parameter --- this choice allows us to compare the case of
interest ($v=0.5$, corresponding to the superexchange limit $\alpha=0$ of the
orbital model) to two limiting cases that are easier to characterize,
and which we address first: (i) $v \gg 1$ and (ii) $|v|\ll 1$.

In case (i), for $t_{c,c'}\equiv 0$ the potential energy is minimized by
$O(2^L)$ degenerate ground states (maximally flippable states \cite{RF5}).
Once quantum fluctuations are turned on via $t_{c,c'}\ne 0$, this
degeneracy is lifted and a particular ordered phase is selected by a
quantum order-by-disorder mechanism;\cite{3MS} for $|t_{6b}|<|t_{6a}|$
one finds\cite{pertv}  a columnar VBC.
In particular, for $t_{6b}=t_{6a}/3=1/4$, this is confirmed by the real-space
correlations depicted in Fig.~\ref{corrd}(a) on a 36-site cluster and at
$v=1.5$; these correlations clearly indicate a columnar pattern.

\begin{figure}
\begin{center}
\includegraphics[width=7.8cm]{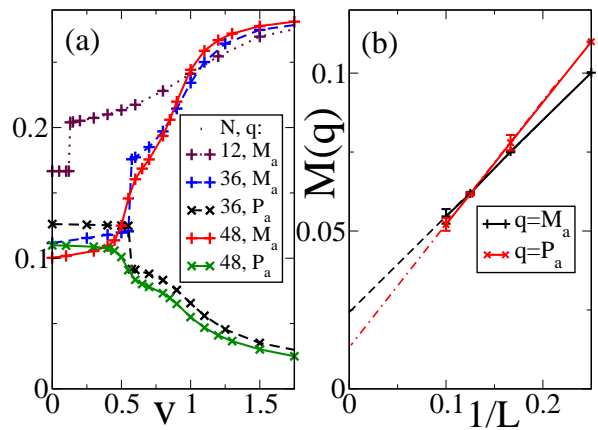}\\
\caption{\label{Gapcorr}(Color online)
  (a) Dimer-dimer order parameter $M(\vec q)$ [Eq.~(\ref{eqstruc})] at
$\vec{q} = P_a$ or $\vec q = M_a$ [see Fig.~\ref{clust}(c)] as
function of $v$ obtained from ED on clusters of $N=12$, $36$ and $48$ sites.
(b) Scaling of $M(\vec q)$ (obtained from GFMC) with inverse linear size
for $v=0$, and either $\vec q = M_a$ (circles) or $\vec q = P_a$ (squares);
dashed lines are a guideline to the eye.}
\end{center}
\end{figure}

More quantitatively, the associated translational symmetry breaking is
reflected by a Bragg peak, at points $\vec q = M_\gamma$ of the Brillouin
zone, in the dimer-dimer order parameter
\begin{eqnarray}
M({\vec q})=\frac{1}{N}
\left\{ \left\langle\Phi_0\left|\sum_{i,j,\gamma}
e^{i{\vec q}\cdot ({\vec r}_i-{\vec r}_j)}d_{i,{\vec{e}_\gamma}}
d_{j,{\vec{e}_\gamma}}\right|\Phi_0\right\rangle\right\}^{1/2}\,,
\label{eqstruc}
\end{eqnarray}
where $|\Phi_0\rangle$ is the ground state for a cluster considered.
Here $d_{i,{\vec{e}_\gamma}}=0$ $(1)$, if a dimer is absent (present)
on the bond $\langle ik\rangle\parallel\gamma$ such that the image of the
site $i$ by the translation of ${\vec{e}_\gamma}$ is $k$. For $v\ge 1.5$
and $N\ge 36$,
$M(M_a)$ exceeds $95\%$ of the maximal value $1/\sqrt{12}$ obtained for a
fluctuation-free columnar order, see Fig.~\ref{Gapcorr}(a); and this order
parameter stays unambiguously finite in the TL for $v \geq 1.0(2)$.

For case (ii), dimer correlations suggest a different translational
symmetry breaking in the TL, shown in Fig.~\ref{corrd}(b,c).
The Bragg peaks in $M(\vec q)$ found at both $M_a$ and $P_a$
[see Fig.~\ref{Gapcorr}(a), and size scaling for $v=0$ in
Fig.~\ref{Gapcorr}(b)] clearly indicate an ordered pattern
invariant by $2\pi/3$-rotations, with a 12-site unit cell
called a plaquette in the following.
This phase has been found to be stabilized by an $l=4$-loop
kinetic term, either isolated \cite{RF5} or along with moderate $l=6$-loop
terms bringing extra resonances within a plaquette.\cite{VR6} When $v$
increases, $M(P_a)$ decreases gradually, in parallel with the increase of
$M(M_a)$, and eventually vanishes in the TL in the range where the large
values of $M(M_a)$ indicate a columnar order, see Fig.~\ref{Gapcorr}(a).
For a value $v \gtrsim 0.5$, one observes in this figure
pronounced jumps in both order parameters, and between this value and $v
\simeq 0.8$ it is not clear whether both quantities, or only $M(P_a)$,
vanish in the TL. This can correspond either to the onset of a RVB spin
liquid \cite{RF5} or to a transition point between two distinct VBCs.
Unfortunately, for $v\gtrsim 0.6$ the GFMC suffers from a lack of
convergence, which restricts the available cluster sizes,
\cite{RF5,VR6} while the correlations obtained by ED are more
consistent with a columnar phase.

\subsection{\label{tgap}Topological gap}

We also discuss the behavior as function of $v$ of another quantity
easily obtained within the effective QDM: the so-called
\textit{topological gap} $\Delta E_t$,\cite{RK88,K61,3MS} which is well
defined a on periodic system, e.g. a torus, for a QDM where off-diagonal
terms correspond only to local updates (this is the case here).
On a torus accommodating the rotation symmetries of the lattice,
it is defined as follows,
\beq
\Delta E_t=\left|E_0(W_c=W_a=1)-E_0(W_c=W_a=-1)\right|,
\label{topo}
\eeq
where $W_\gamma=(-1)^{n_\gamma}$ are topological invariants, such that
$n_\gamma$ is the number of dimers crossing a line parallel to the
$\gamma$ axis and winding around the torus. Since the parity of
 $n_a$ and $n_c$ is conserved by all kinetic terms of the QDM,
each topological sector, or subspace spanned by all dimer configurations with
fixed $W_a$ and $W_c$, is well-defined. We also remark that, on clusters
invariant by point group symmetries of the triangular lattice, the two
sectors such that $W_a+W_c=0$ have the same spectrum as either the $W_a=W_c=-1$
sector, or the $W_a=W_c=1$ one, depending on the cluster size.

The topological gap provides valuable insights into the nature of the
ground state. In a VBC this gap grows as the linear size $L$ of the
system. This is because the ordered pattern fits in only one of the two
topological sectors $(W_a,W_c)=(1,1)$ and $(-1,-1)$; and
 $\Delta E_t$ corresponds to an
excitation disturbing the crystalline order over a whole winding loop.
On the contrary, in a dimer liquid all topological sectors are fully
equivalent in the TL and the topological gap is
simply a finite-size effect, e.g. for a gapped $\mathbb{Z}_2$ liquid
it behaves as $L^\delta e^{-L/\xi}$, with $\xi>0$ and $\delta$
a constant depending on the parity of $L/2$.\cite{RF5}

\begin{figure}[t!]
\begin{center}
\includegraphics[width=7.8cm]{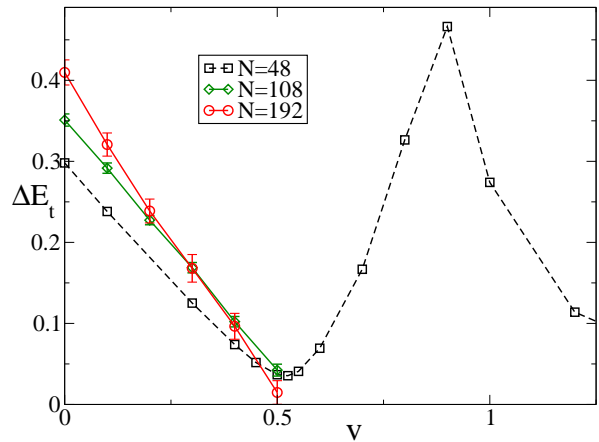}
\caption{\label{gtop} (Color online) Topological gap $\Delta E_t$
[Eq.~(\ref{topo})] versus $v$ obtained for the QDM using large clusters of
$N=3L^2$ sites with periodic boundary conditions. The data are
obtained by exact diagonalization for $L=4$ and with quantum
Monte Carlo for $L\ge 6$.}
\end{center}
\end{figure}

In the present case, on a large cluster of size $N=3L^2$ [see Fig.~\ref{gtop}
with $L\le 8$], $\Delta E_t$ decreases linearly with $v$ from a finite value
at $v=0$ [expected in a plaquette phase where the ground state is in
the sector $(W_c=W_a=(-1)^{L/2})$] down to almost zero for $v_c=0.55(5)$.
When $v$ increases further it takes again a finite value in the
columnar phase: here, the ground state is found in the sector $(W_c=W_a=1)$
(and, if $L/2$ is odd, also in sectors such that $W_c=-W_a$). Thus,
when going from one phase to another,
depending on the parity of $L/2$ the ground state changes or not its
topological sector;
we have verified that the value of $v$ minimizing $\Delta E_t(v)$ is
roughly independent of $L$, so that this minimum is a good indicator
for the transition. In contrast, the maximum of $\Delta E_t(v)$ for
$v\simeq 0.9$ in the $N=48$ case merely signals a crossing between a
non-local excitation of energy $\Delta E_t\propto 1/v^3$
(as expected from perturbation theory in the large $v$ regime)
and another non-local excitation which is of lower energy close to the
transition to the plaquette phase, where this excitation energy
vanishes. From this data the hypothesis of an intermediate RVB phase,
in which $\Delta_t$ extrapolated to the TL would vanish in an finite
range of $v$, seems unlikely. Besides, the vanishing of the topological
gap observed for $v_c \simeq 0.55(5)$ coincides with the changes
occurring in the dimer order parameter.

Although we cannot exclude a liquid phase stabilized in a narrow range
of $v \simeq 0.5$, the most probable scenario is thus a first order
transition between plaquette and columnar phases, occuring at
$v\simeq 0.55(5)$; this implies that the former phase is expected to be
stabilized in the superexchange regime.

\section{\label{direx} The direct exchange regime}

We concentrate now on the regime $\alpha>1/2$ of the model, where direct
exchange is dominant. For $\alpha \to 1$ all other couplings can be
treated as perturbations and, as we did in the previous sections, it is
again possible to derive an effective Hamiltonian which correctly
captures the ground state properties of the microscopic system. In the
following, we report the presence of an exotic ground state
in this regime, and
propose a complete analysis for its characterization.

\subsection{\label{effd} Effective model within the \textit{avoided-blocking} subspace}

\begin{figure}[t!]
\begin{center}
\includegraphics[width=4.5cm]{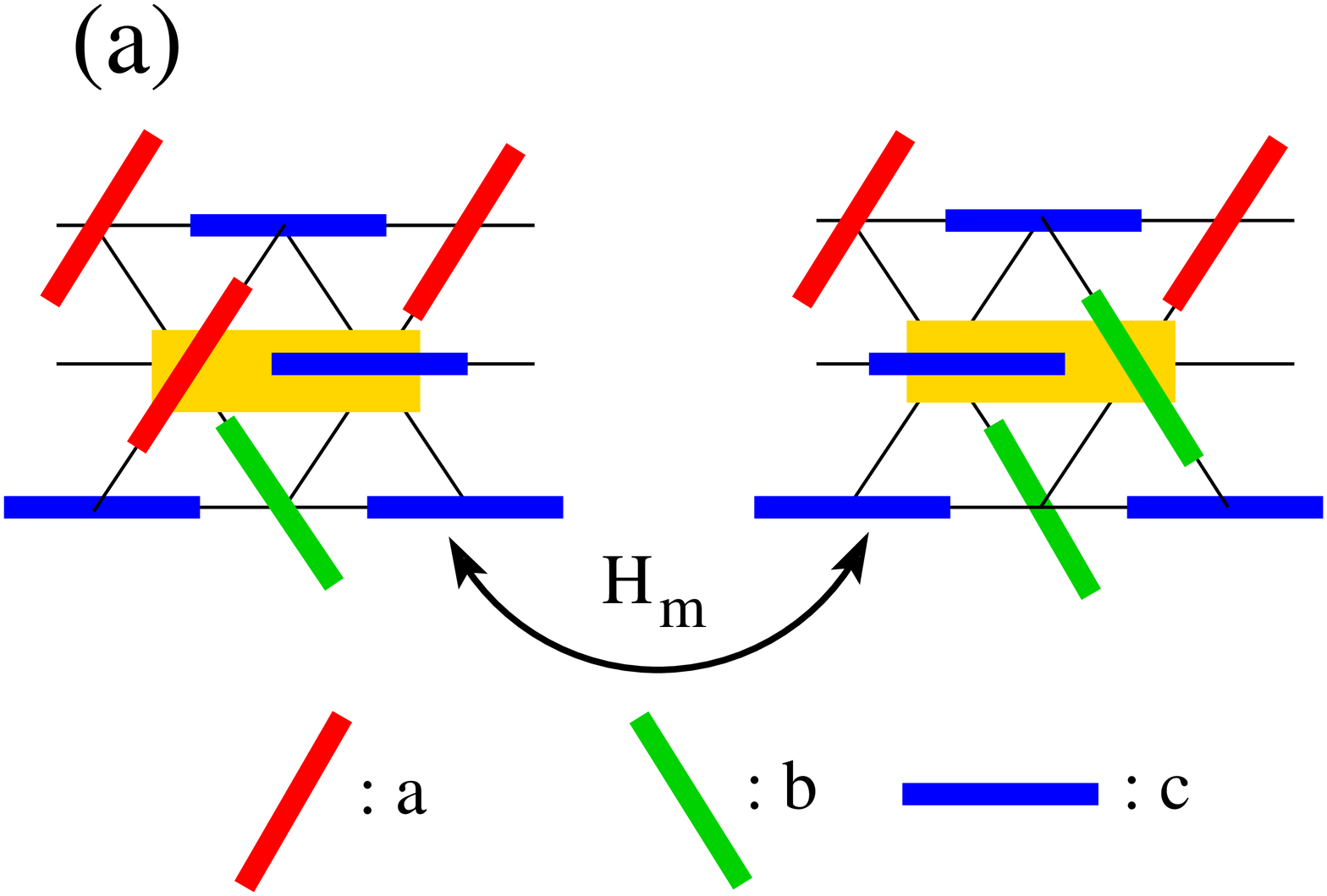}
\hspace{0.2cm}
\includegraphics[width=3.3cm]{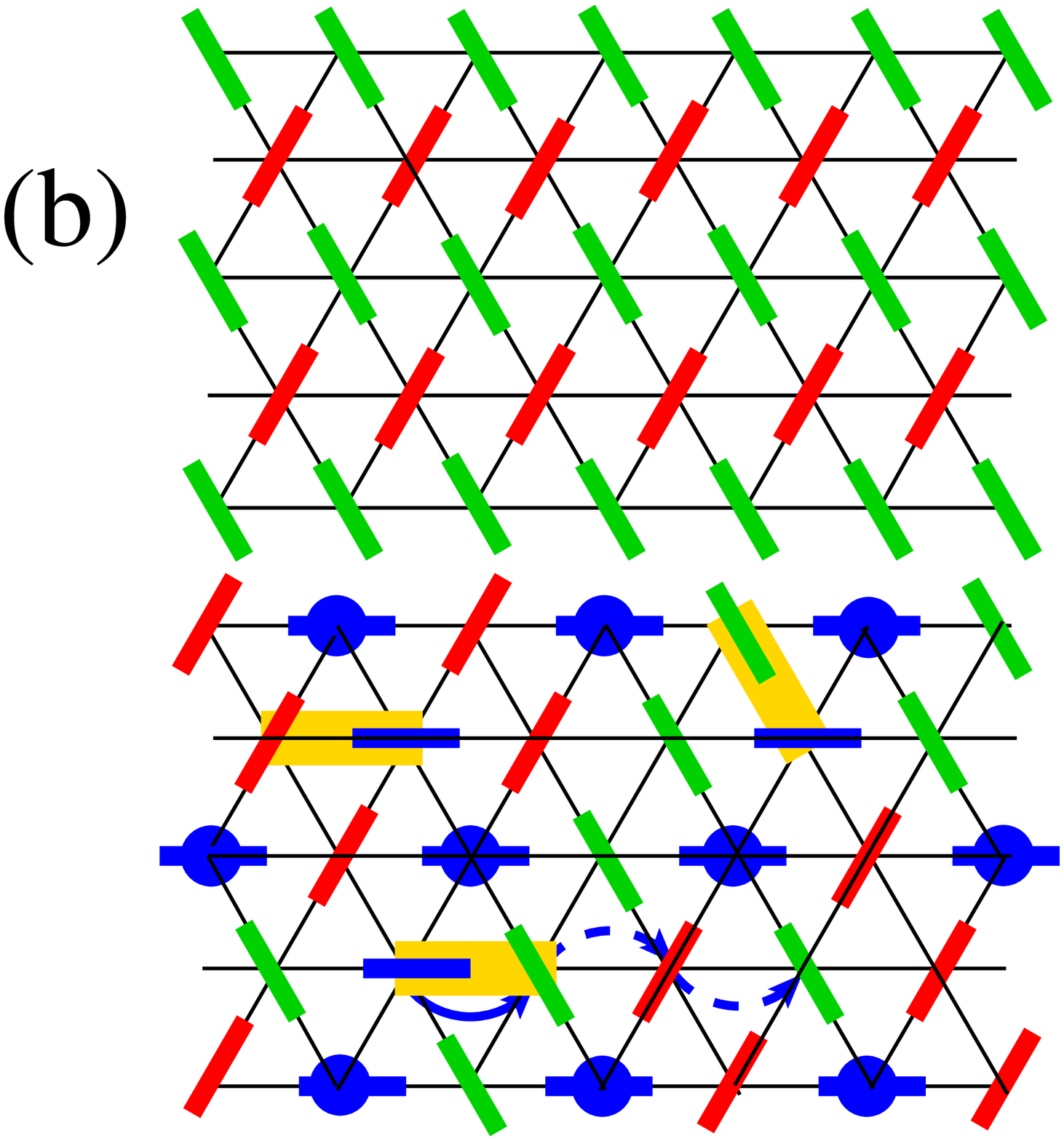}\\
\caption{\label{dirdef}(Color online)
Orbital states for $\alpha\simeq 1$:
(a) typical mixed-exchange process connecting two avoided-blocking
states, which differ on the shaded bond;
(b) ordered patterns ---
the purely collinear pattern (top) is frozen for $\alpha\rightarrow 1$,
while the other (bottom) allows for low-energy dynamics
on the highlighted bonds by consecutive processes triggered by
mixed-exchange terms (arrows), whereas
one sublattice (marked by full dots) is ordered.}
\end{center}
\end{figure}

We start from the limit $\alpha \equiv 1$, in which $\cal{H}$~$\equiv H_d$
is positively definite and selects a macroscopically degenerate ground
state manifold, characterized by $n_{i\gamma}n_{j\gamma}=0$ on every
bond $\langle ij\rangle$ parallel to $\vec{e}_\gamma$. These orbital
configurations, called \textit{avoided-blocking} states, can be
described in a representation where
a rectangle at each site of the triangular lattice represents the
orientation of the occupied orbital at this site ---
a rectangle parallel to $\gamma$-bonds (but centered on the site $i$,
in contrast to dimers of Sec.~\ref{sec:qdm} which were centered on
bonds) means that $n_{i,\gamma}=1$ in the configuration considered
here, see Figs. \ref{dirdef}(a) and \ref{dirdef}(b). The number of
such configurations grows exponentially with system size, i.e., they
are macroscopically degenerate and form a
low-energy subspace separated from higher-energy configurations by a
gap $\simeq 2\alpha \simeq 2$ in this limit.
Close to this limit (for $0<1-\alpha\ll 1$)
the weight of other configurations, with \textit{blocking defects}
--- consisting on bonds $\langle ij \rangle \parallel\gamma$ where on
both sites $i$ and $j$ the $\gamma$ orbital is occupied ---
is proportional to $1-\alpha$.
This scaling can be easily understood within second order perturbation
theory where mixed-exchange terms, which can create/annihilate such
defects, are treated as perturbations w.r.t. direct exchange.

In this regime, the ground state is thus mostly composed of
avoided-blocking configurations, and its properties are determined by
the action of mixed-exchange (and to a lesser extent of superexchange)
perturbations within the avoided-blocking subspace. Indeed, as shown
in Fig.~\ref{dirdef}(a) a single mixed-exchange term can connect
distinct avoided-blocking states, allowing for important quantum
fluctuations. We investigate whether fluctuations can account for a
disordered ground state, or whether the perturbing exchange terms select
a long-range orbital order, by a quantum order-by-disorder mechanism.
To that extent, we define and consider an effective Hamiltonian, found
at first order in a perturbation theory where super- and mixed exchange
are considered as perturbations to the direct exchange Hamiltonian:
\beq
H_{\rm eff} \equiv \sqrt{1 - \alpha}\;\left( \sqrt{1-\alpha}\; \cal{P}
H_{\rm s} \cal {P} +
 \sqrt{\alpha}\; \cal{P} H_{\rm m}\cal{P} \right).
\label{effdir}
\eeq
Here $\cal{P}$ is the projection operator onto the avoided-blocking
subspace, i.e., in the basis of orbital configurations $\{|c\rangle\}$
(eigenstates of all $\gamma_i^{\dagger}\gamma_i^{}$ operators),
${\cal P}$$|c\rangle=|c\rangle$ if $|c\rangle$ is an avoided-blocking
configuration, while $\cal{P}$$|c\rangle=0$ otherwise.

The ground state $|\Psi_0^{\rm eff}\rangle$ of $H_{\rm eff}$ can be
accessed for significantly larger cluster sizes ($N\le 36$ sites in
L\'anczos ED) than the ground state $|\Psi_0\rangle$ of the original
Hamiltonian $\cal{H}$; thus it allows us to address in a more
controlled way the possible existence of long-range order in the TL.

\subsection{Orbital ordering}

\begin{figure}[t!]
\begin{center}
\includegraphics[width=7.8cm]{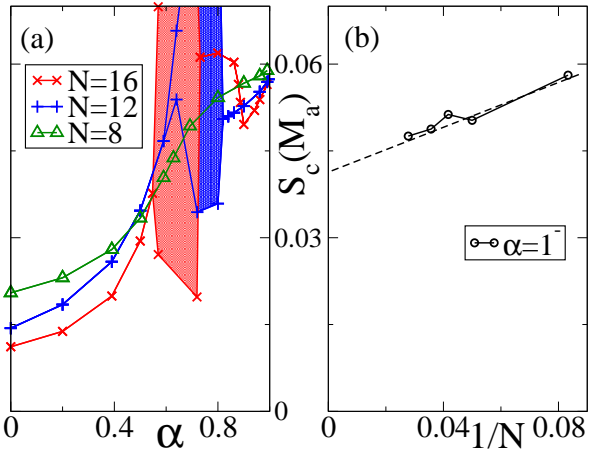}\\
\caption{\label{Sorb}(Color online)
Structure factor $S_c(M_a)$ of orbital correlations, see Eq.(\ref{strucc}),
as obtained (a) within the orbital model of Eq.~(\ref{Hfm}) for the ground
states of $N=8,12,16$ clusters, and for increasing $\alpha$; and (b)
within the effective model of Eq.~(\ref{effdir}) for the
ground states of $N \le 36$ clusters, for $\alpha \to 1$; for computational details see
Ref.~\onlinecite{average}. In (b) the dashed line is a guide to the eyes.}
\end{center}
\end{figure}

The first quantity we consider to address the possibility of orbital
ordering is the structure factor of orbital correlations,
\begin{eqnarray}
S_\gamma(\vec q)=\frac{1}{N^2}\sum_{i,j} e^{i\vec q \cdot
(\vec{r}_i-\vec{r}_j)}
\left\langle \Psi\left| n_{i\gamma}n_{j\gamma}\right|\Psi\right\rangle ,
\label{strucc}
\end{eqnarray}
choosing for $|\Psi\rangle$ either the ground state $|\Psi_0\rangle$ of
the full model or an approximate state $|\Psi_0^{\rm eff}\rangle$ of 
the effective model defined in Sec.~\ref{effd}. Due to
local constraints imposed by dominating direct-exchange interactions,
$S_c (\vec q)$ is expected to be maximal at $M_a$ and $M_b$. The type of
ordered pattern which maximizes this quantity consists of lines with
orbital flavors uniform within a line and alternating between
neighboring lines --- we call this \textit{collinear order}.
On a cluster invariant by point group symmetries, $S_c(M_a)=1/12$ if 
$|\Psi\rangle$ in Eq.~(\ref{strucc}) is chosen as an equal weight
superposition of all such patterns (there are six of them, since lines
parallel to  $\vec{e}_\gamma$ and containing $\gamma$ orbitals are
forbidden here).
Here we find that for $\alpha$ larger than $0.6(1)$, $S_c(M_a)$
takes values of the same order of magnitude, and the comparison
between different clusters [see Fig.~\ref{Sorb}(a)] indicates
that in the TL this quantity may stay finite.

In the limit $\alpha\rightarrow 1$ on which we focus first,
this is confirmed by $S_c(M_a)$ obtained from the ground state
$|\Psi_0^{\rm eff}\rangle$ of the effective model $H_{\rm eff}$, which
for $\alpha \rightarrow 1$ reduces to $H_{\rm m}^{\rm eff}=
H_{\rm eff}(\alpha \rightarrow 1) \propto {\cal P}H_{\rm m} {\cal P}$;
Fig.~\ref{Sorb}(b) clearly indicates that $S_c(M_a)$ computed with
$|\Psi_0^{\rm eff}\rangle$ stays finite in the TL, scaling roughly as
$c_1+c_2/N$ where $c_1>0$. Note that its value for e.g. $N=12$ matches
well the one in Fig.~\ref{Sorb}(a) for the same cluster size and
$\alpha\rightarrow 1$;
besides, both wave functions $|\Psi_0\rangle$ and $|\Psi_0^{\rm eff}\rangle$
have there a large mutual overlap $\simeq 1-k\alpha$. This confirms the
validity of the effective model Eq.~(\ref{effdir}) in this limit, where the
ground state is thus characterized by long-range orbital order inducing a
translational symmetry breaking in the TL.

A collinear phase would be compatible with such a result: indeed the
corresponding ordered patterns [see Fig.~\ref{dirdef}(b)-top] belong to
the avoided-blocking subspace.
Yet, the stabilization of this phase is not expected in
this limit; collinear ordered patterns do not allow for
mixed-exchange fluctuations, whereas these are essential in selecting
a ground state with finite and negative $\langle H_{\rm m}\rangle$ (within the
effective model we estimated $\langle H_{\rm m}\rangle/N\simeq -0.35(5)$
in the TL). With a similar reasoning, one can exclude other candidate
phases with static orbital order.
A dimerized phase could also be considered as a candidate, since
mixed-exchange favors a resonating state on individual bonds, e.g. the
resonating state $(1/\sqrt{2})(|c_i a_j\rangle+|b_i c_j\rangle)$ on a
bond $\langle ij \rangle \parallel c$; yet such a phase,
where blocking defects cannot be avoided, would not be favorable either.

While the nature of the ground state for $\alpha\rightarrow 1$ remains
open after the discussion above,
one notices that an upturn occurs in $S_c(M_a)$ when $\alpha$ is
decreased until reaching a value $\alpha_c\simeq 0.81(1)$
($\alpha_c\simeq 0.86(2)$) for $N=12$ ($N=16$), see Fig.~\ref{Sorb}(a).
This upturn coincides ($N=12$) or not ($N=16$) with the boundary of the
regime with exact twofold ground state degeneracy; it signals a change
in the ground state, which for $\alpha \lesssim \alpha_c$ could be
collinear-ordered. Indeed, the diagonal part of superexchange is
minimized among avoided-blocking states by collinear patterns, in which
e.g. $a$ and $b$ orbitals are never neighboring on a $c$-bond;
far enough from the $\alpha\rightarrow 1$ limit, superexchange becomes
nonnegligible compared to mixed exchange and can favor the collinear order.

\subsection{Insights from low-energy excitations}

\begin{figure}[t!]
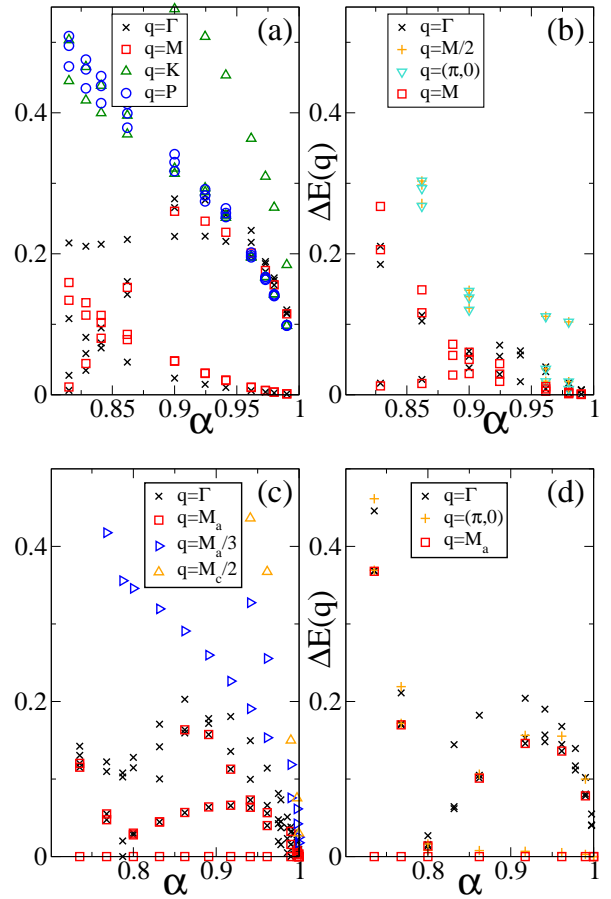

\begin{center}
\includegraphics[width=7.8cm]{fig11a.eps}\\
\vspace{0.4cm}
\includegraphics[width=7.8cm]{fig11b.eps}\\
\caption{\label{spectref} (Color online) Top part --- low-energy spectrum
of the orbital Hamiltonian $\cal{H}$ in the direct-exchange-dominated
regime ($\alpha \ge 0.8$) for clusters of (a) $N=12$ and (b) $N=16$ sites.
Bottom part ---  low-energy spectrum of the effective Hamiltonian
$H_{\rm eff}$ [Eq.~(\ref{effdir})],
for $\alpha \ge 0.7$, and clusters of: (c)  $N=24$, and  (d) $N=16$ sites.}
\end{center}
\end{figure}

We now analyze, still in the direct-exchange-dominated regime, the
low-energy spectra of both the orbital Hamiltonian $\cal{H}$
[Figs.~\ref{spectref}(a) and \ref{spectref}(b)] and the effective
Hamiltonian $H_{\rm eff}$ [Figs. \ref{spectref}(c) and \ref{spectref}(d)].
For $\alpha$ close to $1$, as mentioned in Sec.~\ref{numal}, the lowest
states are found in representations of momenta $\Gamma$ and $M_\gamma$.
\cite{pi0} Above them, one notices excitation branches with energy
\beq
\Delta E(\vec{q}) \simeq c_{N,\vec{q}}\sqrt{1-\alpha},
\label{cnq}
\eeq
i.e., proportional to the amplitude of mixed-exchange processes.
In contrast to low-energy states characteristic of a hypothetical gapped,
orbital ordered phase evoked earlier, these states are not restricted to
high-symmetry momenta, and thus they (or at least some of them) are not
a mere consequence of
translation symmetry breaking in the TL. When comparing between different
sizes the values $\min_{\vec q}\{c_{N,\vec{q}}\}$ corresponding to the lowest of these
branches (which is typically at $\Gamma$ and $M_\gamma$ points) this value
decreases significantly with increasing $N$ and seems to vanish in the TL;
excitations with a similar scaling law [Eq.~(\ref{cnq})] are also found
at other momenta --- they have energies getting smaller with increasing $N$.
These states may follow a certain dispersion law $\omega(\vec{q})$, quantized
here by the finiteness of clusters considered, but corresponding to a
gapless mode of this phase.

This situation contrasts with what we see further away from the
direct-exchange limit, that is, for $0.6 \lesssim \alpha \lesssim 0.8$
within the effective model --- as well as within the full model for the
same range of $\alpha$, see Fig.~\ref{nrj}(b).
In the latter regime, the spectrum is characteristic of the collinear
phase with lowest states \textit{only} at $\Gamma$ and $M_\gamma$,
while excitations with other momenta are found at much higher energies.
Between both regimes, a level crossing occurs between two groups of
quasi-degenerate lowest states, of different nature --- on each side of
the crossing the number and momenta of lowest states indicate the symmetries
broken in the TL. At equal size, this crossing occurs for a value of
$\alpha$ slightly larger in the original model than in the effective
model (e.g. for $N=16$, at $\alpha\simeq 0.9$ and $\simeq 0.8$,
respectively). The origin of this difference in the crossing position is
that the effective model underestimates effects of off-diagonal
superexchange terms which contribute to stabilize the collinear phase.
But the crossing position is not much sensitive to the system size
(within the effective model, on clusters $N= 12, 16, 20, 24$, this
crossing occurs at $\alpha= 0.755(5)$, $0.795(5)$, $0.73(1)$, and
$0.79(1)$ respectively). This shows that the effective model Eq.
(\ref{effdir}) captures well, qualitatively, the transition between
the collinear phase found in a regime where super-, direct and mixed
exchange amplitudes are comparable, and a phase with a distinct,
unconventional orbital order, found in the
direct exchange regime i.e. for $\alpha$ close to 1.

\begin{figure}[t!]
\begin{center}
\includegraphics[width=8.0cm]{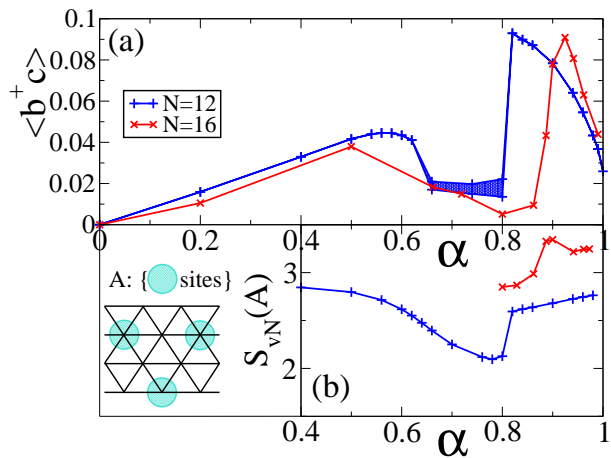}\\
\caption{\label{fluc} (Color online) (a) Amplitude of on-site orbital
fluctuations $\langle b^\dagger c \rangle$ and (b) von Neumann
entanglement entropy ${\cal S}_{vN}(A)$, both evaluated in the ground state of
Eq.~(\ref{Hfm}) for $N=12$, $16$ site clusters \cite{4sub}, as a function of
$\alpha$. The inset shows the subset $A$ of lattice sites used in the
definition of ${\cal S}_{vN}(A)$ [see Eq.~(\ref{vN})].
}
\end{center}
\end{figure}

\subsection{Strong orbital fluctuations in a symmetry-broken phase}

The unconventional structure of the low-energy spectrum observed
for $0<1-\alpha \lesssim 0.2$, with possibly gapless excitations,
contrasts with the --- at least partial --- orbital ordering evidenced
by large values of $S_c(M_a)$. To understand how these features can
coexist, we evaluate, on the ground state of the full model Eq.
(\ref{Hfm}) and for a fixed site $i$, the amplitude of on-site orbital
fluctuations:
\beq
\langle b^\dagger c \rangle = 
\left\langle \Psi_0\left|b^\dagger_i c_i\right|\Psi_0\right\rangle,
\label{eqbc}
\eeq
shown in Fig.~\ref{fluc}(a). This quantity is remarkably large for
$\alpha\geq 0.9$, where it has only weak size-dependence. We checked
that on-site orbital fluctuations have comparable amplitudes for larger
clusters, if computed within the effective model Eq.~(\ref{effdir})
in the same range of $\alpha$. The importance of these fluctuations is
consistent with a finite \textit{orbital compressibility}, i.e., the
fact that the electron density in $\gamma$ orbitals
 can be tuned continuously by an adapted
chemical potential (for details see Appendix B); and with the presumably
gapless excitation modes discussed in the previous paragraph.

When $\alpha$ is decreased further away from the direct-exchange limit,
a drop in $\langle b^\dagger c \rangle$ occurs simultaneously with the
upturn in $S_c(M_a)$ and the crossing in the low-energy spectrum
discussed previously. This is a further indication for the transition
toward static collinear order for $\alpha$ smaller
than this value, while for $\alpha$ close to 1 the mixed-exchange
energy stabilizes a qualitatively more fluctuating ground state. From
the data of Figs.~\ref{nrj}, \ref{Sorb}, and \ref{spectref}, one can
locate the
transition between these two phases at roughly $\alpha \simeq 0.8(1)$.

Another instructive quantity is the von Neumann entanglement entropy
between one sublattice and the rest of the system,\cite{subsvn}
defined as:
\beq
\label{vN}
{\cal S}_{vN}(A) = -{\rm Tr} \left\{\rho_A \ln (\rho_A)\right\},
\eeq
where $\rho_A$ is the reduced density matrix computed for the ground
state of an $N$-site cluster,\cite{4sub} corresponding to a partition
between an ensemble $A$ of $N/4$ sites forming a triangular sublattice
with doubled unit cell [see inset of Fig.~\ref{fluc}(b) or
Fig.~\ref{dirdef}(b)-bottom] and the complementary sites, which form a
kagome lattice. This quantity, shown in Fig.~\ref{fluc}(b), displays a
clear jump at the same position (on the $\alpha$ axis) as the jump in
$\langle b^\dagger c\rangle$ discussed previously. The entropy
${\cal S}_{vN}(A)$ is significantly larger on the right hand side of the
jump, evidencing here a more fluctuating ground state, than on the left
hand side. Besides, we compared this quantity with entanglement
entropies corresponding to other partitions; in those cases, we also
found jumps at the same value of $\alpha$, but with smaller amplitudes
[e.g. when considering ${\cal S}_{vN}(B)$ with $B$ being a
lozenge of 4 sites, on the $N=12$ cluster, between $\alpha=0.8$ and
$\alpha=0.82$ this quantity increases by $\simeq 0.15$ compared to an
increase of $\simeq 0.47$ in ${\cal S}_{vN}(A)$].

These features, in a ground state with spontaneously broken translation
symmetry, motivate us to suggest the following scenario:
the symmetry breaking corresponding to translations by vectors
$\vec{e}_\gamma$ ($\gamma=a,b,c$) allows the ground state to have
a structure where four different sublattices play non-equivalent roles;
in a simplified picture [see Fig.~\ref{dirdef}(b)-bottom] one sublattice
is ferro-orbitally ordered, with e.g. $c$-orbitals covering this
sublattice. On other sites, which alone form a kagome lattice, the
number of electrons of a given orbital flavor fluctuates mainly via
mixed-exchange processes. By these processes, an electron with $c$
orbital flavor can propagate along $c$ bonds as long as it doesn't meet
another electron of this flavor; at each step of this propagation, the
electron moving in the opposite direction converts its orbital flavor
from $a$ to $b$ or vice versa. In other words, such a succession of
mixed-exchange processes on neighboring bonds (see plain and dashed
arrows in Fig.~\ref{dirdef}(b)-bottom) can be seen as the effective
motion, at arbitrary distance and \textit{without direct-exchange-induced
energy cost}, of a single $c$ orbital along a line containing alternating
$a$ and $b$ orbitals. Similar effective single-orbital motions are also
possible in other directions; the low-energy orbital dynamics are thus not
confined in one spatial dimension but rather on an effective kagome lattice.
The freezing of the complementary sublattice may be interpreted as an
order-by-disorder effect. One can also note that the lattice symmetry breaking
evidenced here has been predicted for interacting $t_{2g}$ electrons in such a
geometry, due to the structure of hoppings.\cite{Kos03}

Altogether, this scenario allows
to understand qualitatively that, in the direct-exchange-dominated
regime for $\alpha \gtrsim 0.8$, the ground state is characterized by:
(i) large on-site orbital fluctuations,
(ii) a significantly higher entanglement entropy $S_{vN}(A)$ than in
the collinear state favored for smaller $\alpha$, and
(iii) above the former ground state there may be a continuum of low-energy
excitations, possibly gapless.
This phase would be an orbital analog of a (bosonic) supersolid \cite{HMT}
or (fermionic) pinball liquid,\cite{HF} and is in any case an original and
rather exotic type of ground state in the context of orbital models.

\section{\label{ccl} Summary and Conclusions}

In this paper we have addressed a situation where large Coulomb
interactions lead, even though the spin degrees of freedom are frozen in
a ferromagnetic phase, to a high frustration of effective interactions
between $t_{2g}$ orbital degrees of freedom. This is due to two factors:
first, even when considered separately, both
superexchange or direct exchange interactions are frustrated on the
triangular lattice, while analog models on e.g. a square lattice would
allow for orbitally ordered ground states; second, in the present model
there is a competition between three distinct exchange mechanisms which
naturally coexist in the compounds such as NaTiO$_2$, motivating this
model. To address the nature of the ground states selected by these
interactions, we studied both the orbital model itself and two effective
models adapted to extreme situations. For these studies we employed
numerical techniques --- mostly exact diagonalization, extended by
Monte Carlo techniques for one of these models. This allowed us to
characterize several ground states, depending on the parameter $\alpha$
governing the ratios between amplitudes of the three exchange interactions:

(i) In the regime of dominant superexchange ($\alpha \ll 1$), we found
concordancing evidence for a dimerized phase, where electrons pair into
orbital singlets on nearest neighbor bonds. The effective quantum dimer
model which we derived in this context allowed us to characterize this
phase as a plaquette valence bond crystal (VBC) with a large unit cell
(of 12 sites), and which is stabilized by resonances between singlets
within this unit cell. We also found that slight modifications of
interaction parameters might lead to a transition to another, columnar,
VBC; this suggests that such a system should remain disordered down to
very low temperatures. Yet, the tendency towards dimerization and the
mechanism stabilizing the plaquette VBC seem to be robust to small
mixed exchange interactions perturbing the superexchange ones, for
$\alpha \lesssim 0.6(1)$.

(ii) In the intermediate
regime where superexchange, mixed exchange and direct exchange
interactions have comparable amplitudes, for
$0.6(1)\lesssim\alpha\lesssim 0.8(1)$, we identified a different phase,
with \textit{collinear} orbital ordering in lines, where the orbital
flavor is uniform within a line but alternates between neighboring
lines. This phase, with gapped excitations, seems to be stabilized by
the joined effects of direct exchange and superexchange interactions
--- the former select a large number of low-energy orbital
configurations (\textit{avoided-blocking states}), among which those
favored by superexchange couplings are the collinear patterns.

(iii) Eventually, in the third regime, for $\alpha\gtrsim 0.8(1)$ where
direct exchange is by far the dominant exchange mechanism, we found an
original type of ground state, resulting from the action of subdominant
mixed exchange interactions within avoided-blocking states. This ground
state is characterized by a spontaneous symmetry breaking, which allows
for a structure where orbital long-range order and strong orbital
fluctuations develop on distinct sublattices; the corresponding phase
has low-energy modes, presumably gapless, within the avoided-blocking
subspace. The spatial ground state structure indicated by our results
suggests that it might be an original realization of supersolid- or
pinball-like behavior in orbital physics.

Based on these findings,
several open questions are left for future studies: in the
superexchange-dominated regime, the proximity to the columnar-plaquette
transition within the effective dimer model we considered indicates that
a refined dimer model, including terms found at further order in the
overlap expansion, might predict a columnar ground state, or even a
third dimerized phase, for the superexchange regime. In the
direct-exchange-dominated regime, it would also be valuable to
confirm --- or invalidate --- our scenario by complementary techniques,
ranging from mean field approaches to variational techniques,
applied either to the orbital model or to the effective model adapted
to this regime.
Independently from these issues, one could also investigate whether the
phases we evidence here might be realized in existing compounds --- for
this purpose one should include small octahedral distortions (which
exist in NaTiO$_2$~[\onlinecite{Cla97}]) along with the various exchange
couplings considered here.
An investigation of properties of the resulting orbital model, if
compared to hypothetical (e.g. thermodynamic, or of resonant inelastic
x-ray scattering) measurements on titanates in high magnetic fields,
would allow to estimate the relative amplitudes of the different
exchange mechanisms; one could then consider these amplitudes
(or this value of $\alpha$) and focus on the low-field regime,
to study the possibility of a spin-orbital liquid phase.

\acknowledgments

We thank G. Jackeli and K. Penc for insightful discussions, and J.
Chaloupka for discussions and technical assistance. We acknowledge
support by the European Science Foundation (Highly Frustrated
Magnetism, Grant No. 2525) (F.T.), the French National Research
Agency, Grant No. ANR 2010 BLANC 0406-0 (A.R.), and the Polish
National Science Center (NCN) under Project No. N202 069639 (A.M.O.).
A.R. thanks CIMENT cluster facilities (Grenoble, France) for
computation time.

\appendix

\section{Resonance processes in the superexchange limit}

In the superexchange regime, for $0\le\alpha\ll 1$, the tendency toward
dimerization is quite strong and favors {\it a priori\/} a large number
of nearest neighbor orbital singlet coverings, where a singlet is
created (from the vacuum) by the operator $d_{ij}^{\dagger}$ on a bond
$\langle ij\rangle$
--- for instance, for a bond $\langle ij\rangle \parallel a$,
\beq
d_{ij}^{\dagger}\equiv \frac{1}{\sqrt{2}}\left(b^{\dagger}_i b^{\dagger}_j -
c^{\dagger}_i c^{\dagger}_j\right).
\eeq
Distinct singlet coverings arise naturally due to the form of
$H_{\rm s}$ and are non-orthogonal to one another.
In a dimerized phase, the ground state composed of these coverings may
undergo resonances which lower its
energy and determine the orbital correlations. To quantify the importance of
resonance processes between different singlet coverings we
define the quantity $P_4$ related to resonances on the smallest
non-trivial loop, that is on a lozenge. By labeling sites on a lozenge
in a way shown in Fig.~\ref{ndim}, this quantity reads as:
\beq
P_4=\left\langle\Psi_0\left|d^\dagger_{13}d^\dagger_{24}d_{12}d_{34}+
d^{\dagger}_{12}d^{\dagger}_{34}d_{13}d_{24}\right|\Psi_0\right\rangle,
\eeq
where $|\Psi_0 \rangle$ is the ground state found in exact
diagonalization on the cluster considered. Typical values of $P_4$ are $P_4\simeq
0.29(2)$ in the superexchange limit $\alpha=0$, as seen in Fig.~\ref{ndim}(b),
and decrease with increasing $\alpha$ in the same way as the dimer expectation
value $n_d$. This indicates that, while fluctuations out of the dimerized subspace
increase, the resonance processes stabilizing the dimerized phase subsist
until this phase becomes less favorable than a non-dimerized phase. The values
found on $N=12,16$ clusters can be compared to the extreme (maximal) value
corresponding to a $2\times 2$ cluster, where such a resonance (or flip)
affects the whole system. Assuming here again periodic boundary conditions (so
that all orbital flavors are equivalent), one can easily check that the ground state
of this cluster at $\alpha=0$ reads:
\bea
\left|\Psi_0^{\alpha=0}\right\rangle&=&\frac{1}{\sqrt{6}}\Big\{
  |a_1 a_2 a_3 a_4\rangle
+ |b_1 b_2 b_3 b_4\rangle + |c_1 c_2 c_3 c_4\rangle\Big\} \nn\\
&-&\frac{1}{\sqrt{12}}\Big\{
  |a_1 a_2 b_3 b_4 \rangle + |b_1 b_2 a_3 a_4 \rangle
+ |a_1 c_2 c_3 a_4\rangle \nn\\
&+& |c_1 a_2 a_3 c_4 \rangle
+ |b_1 c_2 b_3 c_4 \rangle + |c_1 b_2 c_3 b_4\rangle\Big\},
\eea
and from this it follows that
\begin{equation}
P_4 = \frac12+\frac{\sqrt{2}}{3} \simeq 0.971\,.
\end{equation}
Such a large value cannot be expected on significantly larger clusters, where
resonances on lozenges coexist with resonances on loops of length $l\ge 6$;
but even there, values $P_4 \simeq 0.29(2)$ for $\alpha\rightarrow 0$ prove
that the contribution of such resonances to orbital dynamics is significant in
this regime, leading to the ground state energy per site $E_0\simeq -1.25(5)$.
Note that this energy is much lower than expected for a static singlet
covering. In fact, the superposition $|\Psi_{\rm var}\rangle$ of columnar
singlet coverings has an energy per site,
\begin{equation}
\langle\Psi_{\rm var}|H_{\rm s}|\Psi_{\rm var}\rangle = -0.5.
\end{equation}

\section{Influence of an external chemical potential in the
  direct-exchange regime}

\begin{figure}[t!]
\vspace{1cm}
\begin{center}
\includegraphics[width=7.8cm]{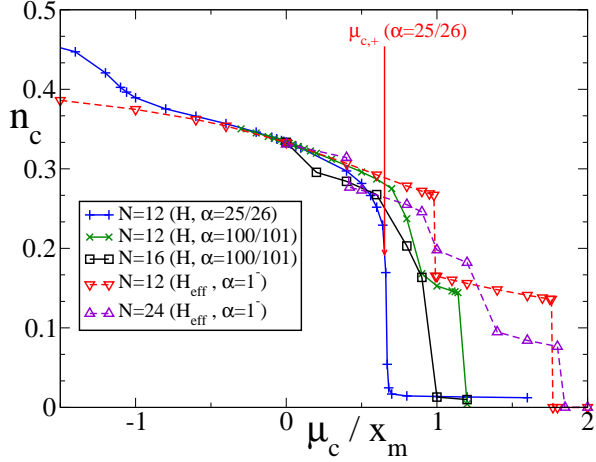}\\
\caption{\label{nmuc} (Color online)
Electron density in $c$ orbitals per site,
$n_c$, as a function of the ratio of $\mu_c$ over mixed-exchange amplitude
$x_m=\sqrt{(1-\alpha)\alpha}$, either within the full model ($\cal{H}$
augmented by the chemical potential term) on periodic clusters ($N=12, 16$
sites) for different $\alpha$ values; or within the constrained model
$H_{\rm eff,c}$ [Eq.~(\ref{effc})],
as obtained for clusters of $N \le 24$ sites for $\alpha \to 1$.
}
\end{center}
\end{figure}

An interesting aspect shedding light onto ground state properties for
$\alpha \rightarrow 1$
is its sensitivity to an external orbital field, or chemical potential,
distinguishing one type of orbital (say, $c$ ones) from the two others.
The effective model augmented by this potential reads now:
\beq
H_{\rm eff ,c}= H_{\rm eff} + \mu_c \sum_i n_{i,c}.
\label{effc}
\eeq
Similarly, within the full orbital model, one can add such a chemical
potential of amplitude $\mu_c$. When this one is positive and large compared
to the mixed-exchange amplitude $x_m=\sqrt{\alpha-\alpha^2}$, the chemical
potential selects, among avoided-blocking configurations, those which minimize
(to zero) the population of $c$ orbitals defined as
\beq
n_c=\frac{1}{N}\sum_i \langle \Psi_0 | n_{i,c}|\Psi_0 \rangle.
\label{eqnc}
\eeq
The selected configurations are two collinear states with alternating rows of
$a$ and $b$ orbitals, see Fig.~\ref{dirdef}(b)-top. For $\mu_c \ll -x_m$
(but both quantities small before the direct-exchange amplitude), the chemical
potential favors the four other collinear states, obtained from the two
previous ones by a lattice rotation combined with a flavor permutation, and
now maximizing $n_c$ within the avoided-blocking subspace. We show in
Fig.~\ref{nmuc} the dependence of $ n_c$ on $\mu_c$ within the constrained
model $H_{\rm eff,c}$ in the direct-exchange limit $\alpha \rightarrow 1$,
as well as within the full orbital model for two values, small but finite,
of $1-\alpha$.

Instead of a direct transition between the two ordered regimes mentioned
above, characterized respectively by $n_c=0$ and $n_c=1/2$ up to quantum
fluctuations, we observe an intermediate regime where $n_c$ varies
continuously with $\mu_c$. In this regime, the ground state has a finite
\textit{orbital compressibility} (defined here as $\frac{dn_c}{d\mu_c}$).
When the mixed-exchange amplitude $x_m$ is small, the values
of the chemical potential corresponding to transitions from the
intermediate regime to both collinear ordered ones, $\mu_{c,\pm}$
[see Fig.~\ref{nmuc} for $\mu_{c,+}$], are close to $\pm x_m$.
When $\alpha$ is decreased, one sees that $\mu_{c,+}$ decreases slightly ---
this behavior is consistent with our indications for a collinear phase when
$\mu_c=0$ and $\alpha \lesssim 0.8$.
Results for the constrained model confirm the presence of the intermediate
regime, with a finite negative slope of $n_c(\mu_c)$ in the vicinity of
$\mu_c=0$, separated from a large-$\mu_c$ regime where $n_c=0$ by a
succession of discontinuities in the range $1 \leq \mu_c/x_m \leq 2  $
(the multiplicity of discontinuities is a finite-size effect). The
finite orbital compressibility at and in the vicinity of $\mu_c=0$,
combined with the long-ranged orbital correlations [evidenced e.g. by
$S_c(M_a)$ in Fig.~\ref{Sorb}] constitute similarities between this
phase and a bosonic supersolid: in both situations, order develops
on one sublattice, while either the number of bosons or the population
of $c$-orbitals can fluctuate on the other sublattices, allowing for a
finite compressibility.


\begin{thebibliography}{99}
\bibitem{3sym} This exact degeneracy originates in a symmetry of the model
  combining a $2\pi/3$ spatial rotation and a permutation of flavor indices;
  there, the ground states are found in representations with character
  $e^{i\pm 2\pi/3}$ for this symmetry.
\bibitem{pertv} Among maximally flippable states, the non-columnar
                ones are destabilized at order $p=3$ or $p=4$ in
                perturbation. This is consistent with a scaling
                $\sim 1/v^3$ for the lowest excitation energies to
                such states, as found on clusters $N=36,48$.
\bibitem{average} For clusters of $N=8$ sites [within the full model of
Eq.~(\ref{Hfm})] and of $N=20,24,28$ sites [within the effective model
of Eq.~(\ref{effdir})], which do not possess $C_{6v}$ point group
symmetries, $S_c(M_a)$ is averaged over the ground states of all
clusters related to each other by these symmetries.
\bibitem{4sub}  For the $N=12$ cluster we computed ${\cal S}_{vN}(A)$ on 
                the lowest state of the representation with $\vec{q}=\Gamma$ 
                and character 1 w.r.t. the rotation-permutation symmetry
                mentioned in Ref.~\onlinecite{3sym}
                --- this allows to get rid of artefacts of the two-fold
                ground state degeneracy for $\alpha\leq 0.81(1)$, while
                the jump observed in Fig.~\ref{fluc}(b) around this
                value reflects well the change of nature of ground
                state in the TL.
\end{thebibliography}
\end{document}